\documentclass[10pt]{article}
\pdfoutput=1

\usepackage{amssymb}
\usepackage{amsmath}
\usepackage{amsthm}
\usepackage{mathtools}
\usepackage{float}
\usepackage{caption}
\usepackage{subcaption}
\usepackage[usenames,dvipsnames]{xcolor}
\usepackage{epsfig}
\usepackage{dcolumn}
\usepackage{tikz}
\usepackage{tikz-cd}
\usetikzlibrary{shapes.geometric,arrows,svg.path}
\usepackage{upgreek}
\usepackage{setspace}
\usepackage{enumitem}
\usepackage{array,multirow,bigdelim,arydshln}
\usepackage{appendix}
\usepackage[export]{adjustbox}
\usepackage{xparse}
\usepackage[utf8]{inputenc}
\usepackage{microtype}
\usepackage{bm}
\usepackage{braket}
\usepackage{dsfont}
\usepackage{nccmath}
\usepackage{scalerel}
\usepackage[dvipsnames]{xcolor}
\definecolor{rindou1}{rgb}{0.4431,0.2862,0.7960}
\definecolor{rindou2}{rgb}{0.0078,0.1215,0.4392}
\definecolor{lapis}{rgb}{0.0.0470,0.2941,0.5568}
\definecolor{emerald}{rgb}{0.31, 0.78, 0.47}
\definecolor{pinegreen}{rgb}{0.0, 0.47, 0.44}
\definecolor{jade}{rgb}{0.0, 0.66, 0.42}
\definecolor{teal}{rgb}{0.0, 0.5, 0.5}
\usepackage{jheppub-mod}

\usepackage{hyperref}
\hypersetup{
	colorlinks=true,
	urlcolor=lapis,
	linkcolor=lapis,
	citecolor=lapis,
	bookmarks=true,
	pdfauthor={Nikhil Kalyanapuram},
	pdftitle={},
	pdfdisplaydoctitle=true,
	pdfstartview=FitH,
	backref=false,
	pagebackref=false
}
\definecolor{orcidlogocol}{HTML}{A6CE39}
\tikzset{
	orcidlogo/.pic={
		\fill[orcidlogocol] svg{M256,128c0,70.7-57.3,128-128,128C57.3,256,0,198.7,0,128C0,57.3,57.3,0,128,0C198.7,0,256,57.3,256,128z};
		\fill[white] svg{M86.3,186.2H70.9V79.1h15.4v48.4V186.2z}
		svg{M108.9,79.1h41.6c39.6,0,57,28.3,57,53.6c0,27.5-21.5,53.6-56.8,53.6h-41.8V79.1z M124.3,172.4h24.5c34.9,0,42.9-26.5,42.9-39.7c0-21.5-13.7-39.7-43.7-39.7h-23.7V172.4z}
		svg{M88.7,56.8c0,5.5-4.5,10.1-10.1,10.1c-5.6,0-10.1-4.6-10.1-10.1c0-5.6,4.5-10.1,10.1-10.1C84.2,46.7,88.7,51.3,88.7,56.8z};
	}
}

\newcommand\orcidicon[1]{\raisebox{.05em}{\href{https://orcid.org/#1}{\mbox{\scalerel*{
				\begin{tikzpicture}[yscale=-1,transform shape]
				\pic{orcidlogo};
				\end{tikzpicture}
			}{|}}}}}
\DeclareMathOperator{\A}{\mathcal{A}}
\DeclareMathOperator{\g}{\gamma}
\newtheorem*{acknowledgements}{Acknowledgements}
\newtheorem*{outline}{Outline}

\allowdisplaybreaks
\graphicspath{{figures/}}
\renewcommand{\thefigure}{\thesection.\arabic{figure}}

\title{{\itshape\LARGE On} Polytopes {\itshape\LARGE and}\hspace{.1em} Generalizations {\itshape\LARGE of the}\hspace{.1em} KLT Relations}
\author{Nikhil Kalyanapuram \orcidicon{0000-0002-0870-1261}}
\affiliation{Department of Physics, Pennsylvania State University, University Park PA 16802, USA}
\affiliation{Institute for Gravitation and the Cosmos, Department of Physics, Pennsylvania State University,University Park, PA 16892, USA}
\emailAdd{nkalyanapuram@psu.edu}

\abstract{%
We combine the technology of the theory of polytopes and twisted intersection theory to derive a large class of double copy relations that generalize the classical relations due to Kawai, Lewellen and Tye (KLT) . To do this, we first study a generalization of the scattering equations of Cachazo, He and Yuan. While the scattering equations were defined on $\mathcal{M}_{0,n}$ - the moduli space of marked Riemann spheres - the new scattering equations are defined on polytopes known as accordiohedra, realized as hyperplane arrangements. These polytopes encode as patterns of intersection the scattering amplitudes of generic scalar theories. The twisted period relations of such intersection numbers provide a vast generalization of the KLT relations. Differential forms dual to the bounded chambers of the hyperplane arrangements furnish a natural generalization of the Bern-Carrasco-Johansson (BCJ) basis, the number of which can be determined by counting the number of solutions of the generalized scattering equations. In this work the focus is on a generalization of the BCJ expansion to generic scalar theories, although we use the labels KLT and BCJ interchangeably.
}

\begin{document}

\maketitle
\setcounter{page}{2}

\vfill

\pagebreak

\section{Introduction}\label{sec:intro}
A double copy relation broadly defined is a method by which the scattering amplitudes of a given theory - such as Einstein gravity - can be defined equivalently as a convolution of two scattering amplitudes of some other quantum field theory, such as Yang-Mills. These relations often generate considerable interest due to the perspective they provide on the structure of gravitational amplitudes. The most famous double copy relation is that of Kawai, Lewellen and Tye \cite{KAWAI19861}, which is a statement about the relationship borne by gravity amplitudes to gluon amplitudes at tree level. The relation is expressed by the formula,

\begin{equation}
    \mathcal{A}_{GR}(1,..,n) = \sum_{\alpha,\beta}\mathcal{A}_{YM}(\alpha)S[\alpha|\beta]\A_{YM}(\beta).
\end{equation}
The notation is by now quite standard, but we will review it for completeness. $\A_{GR}$ and $\A_{YM}$ are the scattering amplitudes for gravity and Yang-Mills at tree level, here evaluated for $n$ particles. The Yang-Mills amplitudes are colour ordered, denoted by $\alpha$ and $\beta$. The sum is over $(n-3)!$ colour orderings, known now as a Bern-Carrasco-Johansson basis \cite{Bern:2005hs}. The matrix $S[\alpha|\beta]$ is the KLT kernel, which fuses together the gauge theory amplitudes to return the gravity amplitude. 

One of the most interesting lines of research in the modern study of scattering amplitudes deals with the subtleties of the KLT kernel, which enjoys several interesting mathematical properties. The kernel can be computed as the inverse of a matrix whose entries are $m(\alpha|\beta)$, the double-partial ordered amplitudes of a theory known as biadjoint scalar field theory. This fact was originally discovered by Cachazo, He and Yuan \cite{Cachazo:2013gna}, and is most easily derived in their scattering equations formalism \cite{Cachazo:2013hca,Cachazo:2013iaa,Cachazo:2013iea,Cachazo:2014nsa,Cachazo:2014xea,Cachazo:2015aol}. Indeed, a particular upshot of this formalism is the trivialization of the KLT relations as a simple fact of linear algebra. This formalism, now known as the CHY formalism, inspired a growing understanding of scattering amplitudes in quantum field theories as essentially geometric and topological objects. Two key aspects in this study are the theory of positive geometries and polytopes and the theory of twisted intersection numbers. 

The positive geometry program, which found its inception in the work of Arkani-Hamed \emph{et al.} \cite{Arkani-Hamed:2017mur,Arkani-Hamed:2017tmz} provided a new understanding of scattering amplitudes in biadjoint theories as the volume of a fairly special geometric object, known as the associahedron. In more technical terms, the associahedron $\mathcal{A}_{n}$ of dimension $(n-3)$ admits a convex embedding in the space $\mathcal{K}_{n}$ of Mandelstam invariants for $n$ massless particles. A differential form $\Omega$ of top degree, known in the literature as a canonical form, is fixed uniquely upto a prefactor by simply demanding a logarithmic divergence as the boundaries of the associahedron are approached. The residue of this form on the associahedron can be shown to be equal to the scattering amplitude of the $\phi^3$ at tree level,

\begin{equation}
    \mathrm{Res}_{\mathcal{A}_{n}}\Omega_{n} = m_{\phi^3}(1,...,n).
\end{equation}
The geometric structure of the associahedron and the analytic structure of the canonical form trivialise quantitative features of the scattering amplitude, such as locality and unitarity. The local structure of the amplitude is fixed by the logarithmic divergence condition, while unitarity is assured at tree level by the boundary structure of the associahedron. Any boundary $\mathcal{B}$ of the associahedron factors into two lower point associahedra $\mathcal{A}_{n_1}\times \mathcal{A}_{n_2}$. Approaching the boundary amounts to a unitarity cut, and the residue of the canonical form factorizes accordingly, encoding the factorization of the scattering amplitude, implying unitarity. A polytope equipped with a canonical form is known as a \emph{positive geometry}.

The associahedron encodes the amplitudes of theories with cubic vertices due to its simple combinatorial definition. The vertices of an associahedron $\mathcal{A}_{n}$ are in one to one correspondence with the triangulations of an $n$-gon. A given triangulation of an $n$-gon can be dualized into a trivalent tree, which can be interpreted as a Feynman diagram for a cubic theory. Accordingly, it becomes intuitively clear why the associahedron encodes so naturally the analytic structure of scattering amplitudes for such theories. 

This approach to understanding scattering amplitudes has enjoyed quick progress over the last few years. In particular, much work has gone into trying to extend these ideas to theories which have more complicated interactions. The first in this line of work was due to Banerjee, Laddha and Raman \cite{Banerjee:2018tun}, in which it was found that a class of polytopes known as Stokes polytopes furnish the positive geometries of quartic scalar theories at tree level in the planar limit. Stokes polytopes are convex polytopes whose vertices are in one to one correspondence with \emph{quadrangulations} of an $n$-gon. They differ qualitatively from associahedra in that they are not unique for a given number of particles interacting. Even for $8$ particles, there are two different types of Stokes polytopes, which give rise to partial amplitudes. The total amplitude is obtained by a sum over these partial amplitudes weighted by factors determined entirely by the combinatorics of the Stokes polytopes. This picture was extended to $\phi^p$ theories and to theories with generic interactions (see respectively \cite{Raman:2019utu} and \cite{Jagadale:2019byr}), and it was found that for all these theories, a large variety of polytopes collectively known as \emph{accordiohedra} played the role of the positive geometry. The accordiohedra for a large class of scalar theories at tree level have now been determined, studies of which have led to a renewed interest in their geometric and analytic structures \cite{Jagadale:2020qfa,Kojima:2020tox,John:2020jww}.

Accompanying this line of work nearly simultaneously was the application of a novel class of mathematical objects known as twisted intersection numbers to the study of scattering amplitudes. Originally, the techniques of twisted intersection theory were used to study the KLT kernel in string theory \cite{Mizera:2017cqs}, but have since been applied to general quantum field theories \cite{Mizera:2017rqa}, Feynman integrals \cite{Mastrolia:2018uzb,Frellesvig:2019kgj,Frellesvig:2019uqt,Frellesvig:2020qot} and other aspects of scattering amplitudes, including recursion \cite{Mizera:2019gea} and the double copy \cite{Mizera:2019blq}. 

The essential data characterizing a twisted intersection theory is a one-form $\omega$ on $\mathbb{CP}^{n}$ with a collection of hyperplanes $\lbrace{H_{1},...,H_{N}\rbrace}$ removed, with the property that it diverges logarithmically as the hyperplanes are approached. The residue prescriptions near these singularities encode kinematical information. Defining a quantum field theory scattering amplitude now requires the definition of two forms $\phi_L$ and $\phi_R$, which have middle dimension (real dimension $n$). These forms however must be defined up to a cohomology prescription, where the cohomology is now \emph{twisted} by redefining the exterior derivative as $\nabla_{\omega} = d + \omega \wedge$. Now it can be shown that a pairing can be defined between two such forms. For suitable choices of the forms and hyperplane arrangement, the scattering amplitudes for an enormous class of quantum field theories can be obtained. Indeed, the CHY formula is a special case of a twisted intersection number, in which the hyperplane arrangement yields the moduli space of marked Riemann spheres. For an extensive review of methods and formalism, the reader should consult \cite{Mizera:2019gea}.

One of the advantages of the method of twisted intersection numbers is the trivialization of identities such as the KLT relations, recasting them as statements of linear algebra. These relations, known as the twisted Riemann period relations \cite{cho1995}, can be stated as follows. Suppose we have two forms $\varphi_1$ and $\varphi_2$ belonging to the $n$th cohomology of $\nabla_{\omega}$ and bases $\lbrace{\phi_{i}\rbrace}$ and $\lbrace{\phi'_{i}\rbrace}$ of this cohomology group. The twisted intersection number $\langle{\varphi_1,\varphi_2\rangle}$ can be shown to satisfy,

\begin{equation}\label{eq1.3}
    \langle{\varphi_1,\varphi_2\rangle} = \langle{\varphi_1,\phi_i\rangle}(\mathbf{C}^{-1})^{T}_{ij}\langle{\phi_j,\varphi_2\rangle},
\end{equation}
where the matrix $\mathbf{C}_{ij}$ is the intersection matrix $\langle{\phi_i,\phi'_j\rangle}$. Indeed, when the cohomology is defined on $\mathcal{M}_{0,n}$ and the basis of the cohomology class is chosen to be a Parke-Taylor basis, the intersection matrix simply becomes the KLT kernel in field theory. 

Now, given these two geometric approaches to scattering amplitudes, a natural question to ask is whether or not the advantages of twisted intersection theory can be exploited by the technology of polytopes. This question was addressed by the author for the case of $\phi^4$ theory and Stokes polytopes in \cite{Kalyanapuram:2019nnf}, where the planar amplitudes in this theory were written as intersection numbers, enabling an extension to generic kinematics as well. The extension to all scalar theories was then carried out in \cite{Kalyanapuram:2020tsr}, with Jha. Using this framework, the goal of the present work will be to carry out a systematic study of the scattering equations for generic scalar theories and provide a concrete realization of the twisted period relations for the relevant polytopes as well. 

In order to study the twisted period relations for the polytopes controlling generic scalar theories, known as accordiohedra, some relevant analytic data must first be secured. In particular, the specific convex realizations of accordiohedra in $\mathbb{CP}^{n}$ have to be understood. The twist for accordiohedra, encoding kinematical information now play the role of the scattering equations, the number of solutions of which determines the size of the KLT matrix. The task of finding the KLT relations then devolves upon the definition of a suitable basis of the cohomology and the computation of the intersection matrix of these elements. To ensure that the context and motivations for these calculations are clear, we carry out the analysis for most of the examples worked out in the literature so far, reviewing most of the background material needed for the study.

Throughout the paper, we will only be dealing with scalar theories and won't concern ourselves with theories containing nontrivial kinematical numerators. It is important to note then that what we are really doing is \emph{expanding} scalar amplitudes into a so-called Bern-Carrasco-Johansson basis. For the purposes of simplicity, we will continue to refer to the expansions we will derive as KLT relations, but it should be kept in mind that we are not yet dealing with theories with numerators.

\pagebreak

\begin{outline}\normalfont
In Section \ref{sec:scattering}, we review the construction of accordiohedra as hyperplane arrangements in complex projective space. We explain how these polytopes encode kinematical data by associating to these objects twist differential forms, which specify a cohomology for each polytope. The dimensions of these cohomology groups are computed by finding the number of zeroes of the twist, in the process providing generalizations of the scattering equations of Cachazo, He and Yuan. 

In Section \ref{sec:relations}, we use the data derived in the Section \ref{sec:scattering} to study the \emph{twisted Riemann period relations} for the intersection numbers of accordiohedra, focusing on one and two dimensional examples.

In Section \ref{sec:conclusion}, we conclude the paper by comparing and contrasting the results derived here with the traditional KLT construction and lay out a number of directions for future research.
\end{outline}

\vfill

\begin{acknowledgements}\normalfont
The author thanks Jacob Bourjaily for his helpful comments on the draft and his continued interest and encouragement. The author also thanks Nima Arkani-Hamed, Benjamin de Bruyne, Freddy Cachazo, Simon Caron-Huot, Alfredo Guevara, Raghav Govind Jha, Sebastian Mizera, Seyed Faroogh Moosavian, Andrzej Pokraka and Jaroslav Trnka for comments on the draft and discussions. This project has been supported by an ERC Starting Grant (No. 757978) and a grant from the Villum Fonden (No. 15369).
\end{acknowledgements}

\pagebreak

\section{Accordiohedra and Scattering Equations}\label{sec:scattering}

In this section, we will present a generalization of the scattering equation formalism relevant to the study of the amplitudes defined by accordiohedra. Much of the technical material of this section has already appeared in previous works by the author (see \cite{Kalyanapuram:2019nnf} for applications to Stokes polytopes and \cite{Kalyanapuram:2020vil,Kalyanapuram:2020tsr} for elementary applications to accordiohedra). Accordingly, the main goal of this section will be to provide a number of examples which have not yet been considered or only received treatment in passing, providing a coherent synthesis of existing techniques and setting the context for later sections.

\subsection{Accordiohedra as Hyperplane Arrangements}

\begin{figure}[h]
\centering
\includegraphics[width=0.5\textwidth]{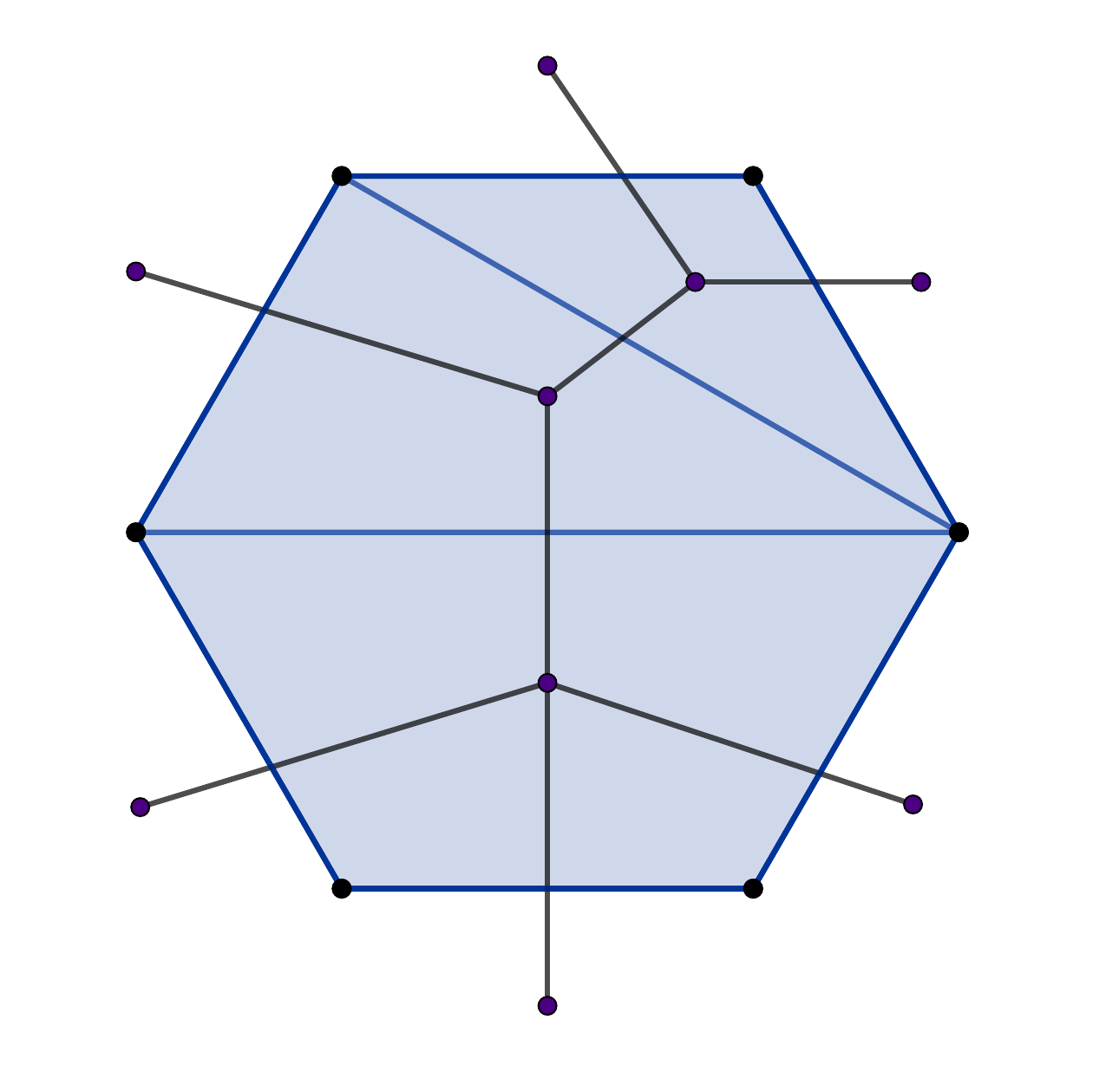}
\caption{Dissection $(13,36)$ of a hexagon and the dual Feynman diagram}\label{fig2.1}
\end{figure}

Accordiohedra are polytopes whose vertices are defined in one to one correspondence with given dissections of an $n$-gon. Such dissections can be mapped directly into Feynman diagrams of a corresponding quantum field theory. The canonical example considered in the literature is the case of a hexagon dissected into two triangulations and one quadrangulation. In the language of quantum field theory, such dissections are exactly dual to tree level Feynman diagrams proportional to $\lambda^2_3\lambda_4$ in a theory with interactions $\lambda_3\phi^3 + \lambda_4\phi^4$. In figure \ref{fig2.1}, this has been illustrated for the dissection $(13,36)$ (vertices are labelled starting at the top left clockwise).

Given a collection of dissections of a given type, accordiohedra are defined for each dissection separately, with the vertices being labelled by those dissections satisfying a compatibility condition with the reference dissection. For a review of the details involved and a number of examples, the reader may consult \cite{Jagadale:2019byr}. This procedure usually results in several different types of accordiohedra for a given class of scattering amplitudes, which must be appropriately summed over in order to produce the correct scattering amplitude.

One of the more mundane reasons why one would expect this to be the case can be realized by a simple counting experiment. Consider the $6$-particle scattering amplitude in $\phi^4$ theory in the planar limit. Three possible scattering channels contribute to this process, namely $X_{14}$, $X_{25}$ and $X_{36}$\footnote{The notation used here is $X_{ij} = (p_{i}+...+p_{j-1})^2$.}. Since the process involves a single propagator, the positive geometry is expected to be one dimensional. There exists no one dimensional polytope with three vertices. Accordingly, it is more likely that specific subsets of the diagrams can be arranged into positive geometries, the sum over which would be expected to yield the amplitude. The three scattering channels each contribute one Stokes polytope. Each of these is a line. These are illustrated in the figure below.

\begin{figure}[h]
\centering
\includegraphics[width=1.1\textwidth]{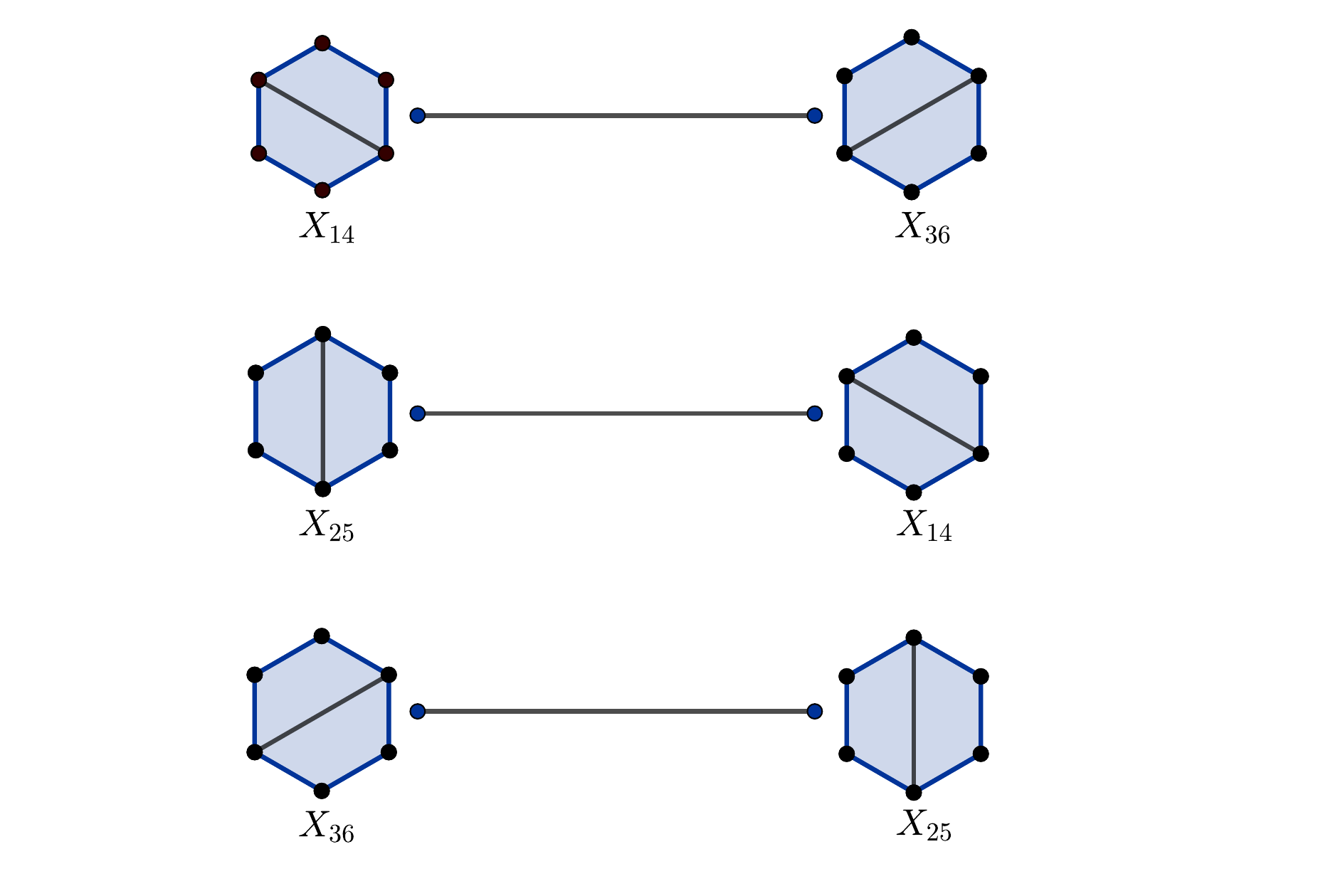}
\caption{Accordiohedra for $6$-particle scattering in $\phi^4$. Reference dissections are on the left.}\label{fig2.2}
\end{figure}

While the foregoing case furnishes the simplest of examples, accordiohedra can be computed for arbitrarily complicated scattering processes. More to the point, since they are convex polytopes (see \cite{Thibault:2017nnf} for more details), they admit realizations as hyperplane arrangements. Correspondingly, twisted intersection theory may naturally be applied to their study. In order to see how this works out for the $6$-particle case, we note that each of the accordiohedra can be realized as a line in $\mathbb{CP}^{1}$ by removing two distinct points $p_1$ and $p_2$. Indeed, this is a hyperplane arrangement in one complex dimension. 

Given a hyperplane arrangement $\lbrace{f_1,...,f_{N}\rbrace}$ in $\mathbb{CP}^n$ realizing an accordiohedron, kinematical data is introduced into the discussion by defining a twist,

\begin{equation}
    \omega = X_{f_{1}}d\ln f_{1}+\dots+X_{f_{N}}d\ln f_{N}.
\end{equation}
The coefficients $X_{f_{i}}$ are the planar $X_{ij}$ variables associated to the facet described by the hyperplanes $f_{i}$.We note parenthetically that the twisted differential $\nabla_{\omega} = d + \omega\wedge$ gives rise to twisted cohomology classes, the homology duals of which can be recovered by the Poincare duality. We will have occasion to use this fact soon.

The twist gives rise to a natural generalization of the scattering equations. Given the foregoing twist, we have,

\begin{equation}
    \omega  = \omega_{x_1}dx_{1}+...+\omega_{x_n}dx_{n}
\end{equation}
where $dx_{i}$ are the the inhomogeneous coordinates on $\mathbb{CP}^{n}$. The twist equations $\omega_{x_i}$ furnish the analogue of the scattering equations for accordiohedra. To make contact with the original idea of the scattering equations, we note that the scattering equations are essentially the twist equations of associahedra, a fact proven in \cite{Mizera:2017cqs,Mizera:2017rqa}. 

With these basic points noted, our purpose in the rest of this section will be to write down the scattering equations for a number of accordiohedra and compute the number of solutions for each system. The data provided in this section will then be used to generalize the KLT relations to accordiohedra in the next section.

\subsection{One Dimensional Polytopes}
\paragraph{$6$ Particles in $\phi^4$ Theory}
For the case of $6$-particle scattering in $\phi^4$ theory, there is a single reference dissection, namely $(14)$, meaning all other dissections can be obtained by cyclic permutations thereof. Accordingly, we have a single accordiohedron to think of - the line with vertices labelled by $(14)$ and $(36)$. This can be realized as the arrangement $\mathcal{A}_{(14)} = \mathbb{CP}^{1}-\lbrace{0,1\rbrace}$ with the twist, 

\begin{equation}
    \omega_{14} = \left(\frac{X_{14}}{x}-\frac{X_{36}}{1-x}\right)dx.
\end{equation}
The scattering equation,

\begin{equation}
    \frac{X_{14}}{x}-\frac{X_{36}}{1-x} = 0
\end{equation}
has a single solution. This may be seen from the fact that this arrangement has one bounded chamber, namely the line between $0$ and $1$. We remark here that there is a hyperplane at $x=\infty$ as well, with residue $-X_{14}-X_{36}$. This point will come up when we discuss the KLT relations in later sections. Finally, we note,

\begin{equation}
    \mathrm{dim}H^{1}_{\omega_{(14)}}(\mathcal{A}_{14}) = 1.
\end{equation}
Indeed, we have the same result for the dissections $(25)$ and $(36)$.

\paragraph{$5$ Particles in $\phi^3+\phi^4$ Theory}
We now consider $5$-particle scattering in $\phi^3+\phi^4$ theory with amplitudes having one cubic and one quartic vertex. At the planar level, there are five diagrams, as shown below

\begin{figure}[H]
\centering
\includegraphics[width=1\textwidth]{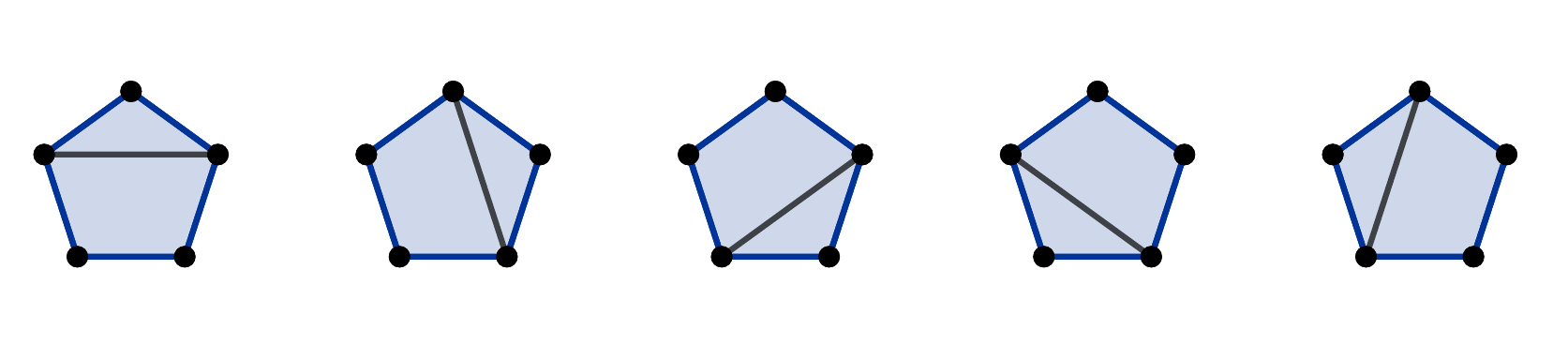}
\caption{Planar Feynman Diagrams for $5$ Particles in $\phi^3 + \phi^4$ Theory}\label{fig2.3}
\end{figure}
In the figure, the diagram to the left is $(13)$, with the rest obtained by cyclic rotations. As a result, we have here only one accordiohedron, corresponding to the reference $(13)$, from which the others may be easily inferred. This turns out to be the line, with the compatible dissections being $(13)$ and $(26)$, which can be realized again by $\mathcal{A}_{(13)} = \mathbb{CP}^{1}-\lbrace{0,1\rbrace}$ with the twist,

\begin{equation}
    \omega_{(13)} = \left(\frac{X_{13}}{x}-\frac{X_{26}}{1-x}\right)dx.
\end{equation}
Once again, we see that the scattering equation has a single solution. Thus we have,

\begin{equation}
    \mathrm{dim}H^{1}_{\omega_{(13)}}(\mathcal{A}_{13}) = 1.
\end{equation}
It goes without saying that this result holds for the remaining dissections as well.

\subsection{Two Dimensional Polytopes}
\paragraph{$8$ Particles in $\phi^4$ Theory}
We turn now to our first nontrivial example. In $\phi^4$, for $8$-particle scattering, we need to consider quadrangulated octagons. There are two inequivalent classes of such quadrangulations, given in the diagrams below.

\begin{figure}[H]
\centering
\includegraphics[width=0.4\textwidth]{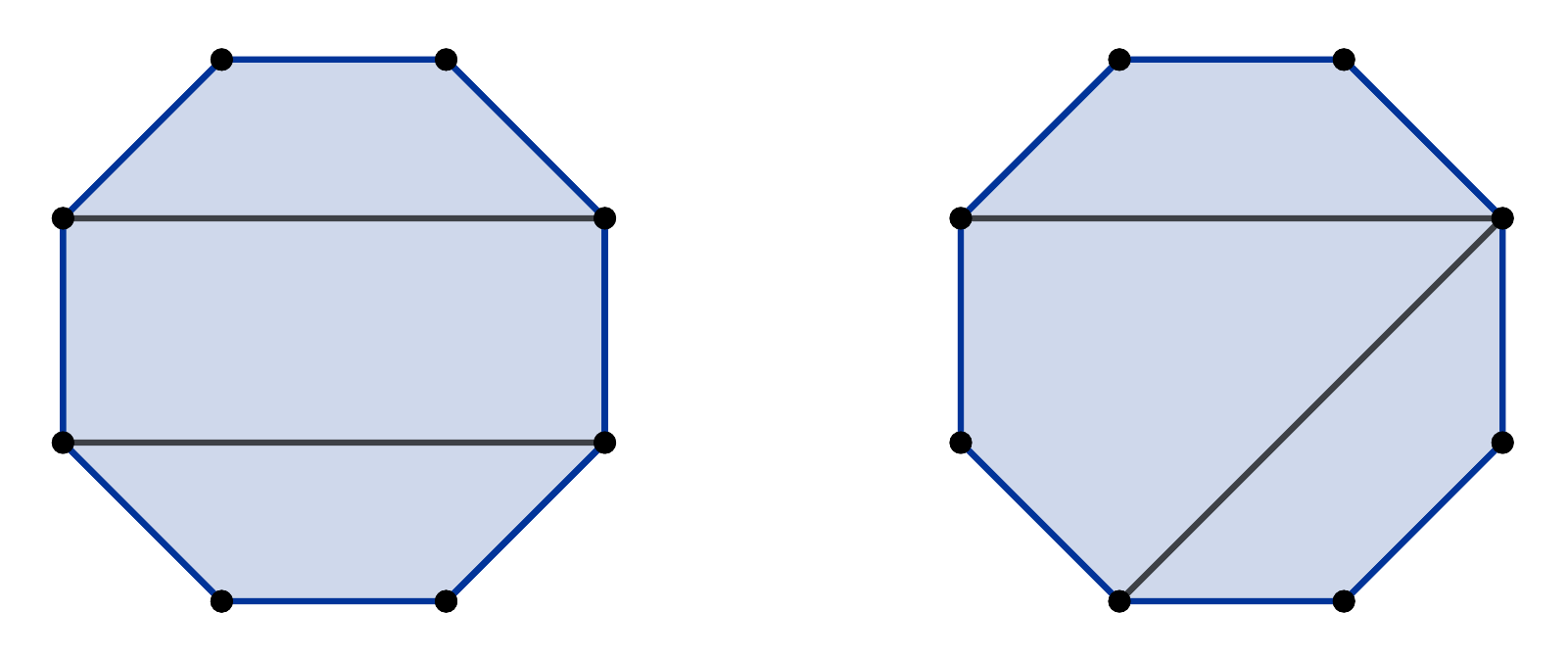}
\caption{Inequivalent Diagrams for $8$-Particle Scattering in $\phi^4$}\label{fig2.4}
\end{figure}
The diagram on the left is dissection $(14,58)$ and the diagram on the right is $(14,47)$. It can be shown that all other planar diagrams for this process can be obtained by cyclic permutations of one of these two diagrams. Now, we need the accordiohedra corresponding to these two different dissections. Let us start with $(14,58)$. The accordiohedron for this reference dissection is a square, as there are four quadrangulations of the $8$-gon which are compatible with this dissection. These are $(14,58)$, $(14,47)$, $(38,58)$ and $(38,47)$. 

\begin{figure}[H]
\centering
\includegraphics[width=0.7\textwidth]{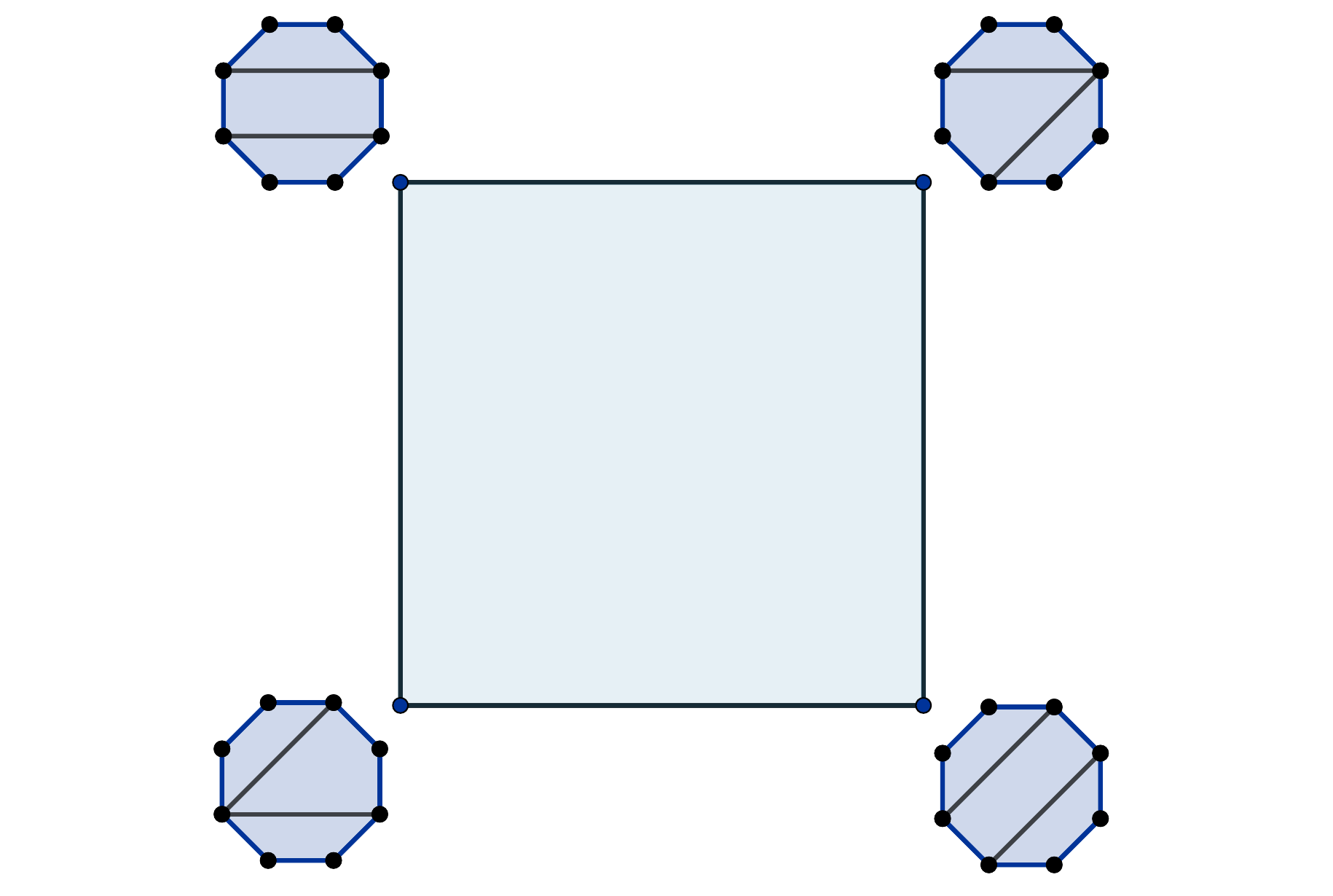}
\caption{Accordiohedron for Dissection $(14,58)$}\label{fig2.5}
\end{figure}
In figure \ref{fig2.5}, the accordiohedron has been represented in terms of the compatible dissections. The upper left vertex in the diagram is the dissection $(14,58)$. As a convex polytope, this can be described by the following collection of hyperplanes,

\begin{equation}\label{eq2.8}
    \begin{aligned}
    f_{1} &= 1-y\\
    f_{2} &= 1-x\\
    f_{3} &= 1+y\\
    f_{4} &= 1+x
    \end{aligned}
\end{equation}
in $\mathbb{CP}^{2}$. Now we come across the first nontrivial case in which the twist has to be properly defined. The facets have to be brought in correspondence with the correct factorization channels. From diagram \ref{fig2.5}, we can identify facets $f_{1}$ through $f_{4}$ with channels $X_{14}$, $X_{47}$, $X_{38}$ and $X_{58}$ respectively, with the twist defined accordingly,

\begin{equation}
    \omega_{(14,58)} = X_{14}d\ln f_{1} + X_{47}d\ln f_{2} + X_{38}d\ln f_{3} + X_{58}d\ln f_{4}.
\end{equation}

The scattering equations can now be read off from the twist. We obtain 

\begin{equation}
    \frac{X_{14}}{1-y}-\frac{X_{38}}{1+y} = 0
\end{equation}
and
\begin{equation}
    \frac{X_{47}}{1-x}-\frac{X_{58}}{1+x} = 0.
\end{equation}

From the above equations, we can infer that again, we have a single solution to the scattering equations. We can record this observation according to the formula

\begin{equation}
    \mathrm{dim}H^{2}_{\omega_{14,58}}(\mathcal{A}_{(14,58)}) = 1.
\end{equation}
This is not especially surprising; a square is after all the product of two lines, each of which has one solution. We remark parenthetically that this phenomenon more generally can be stated as follows. Given a $d = d_1+d_2$ dimensional accordiohedron $\mathcal{A}_{D}$ factorizing into two accordiohedra $\mathcal{A}_{D_1}$ and $\mathcal{A}_{D_2}$ of dimensions $d_1$ and $d_2$ respectively, we have the formula,

\begin{equation}
    \mathrm{dim}H^{d}_{\omega_{D}}(\mathcal{A}_{D}) = \mathrm{dim}H^{d_1}_{\omega_{D_1}}(\mathcal{A}_{D_1})\times \mathrm{dim}H^{d_2}_{\omega_{D_2}}(\mathcal{A}_{D_2}) 
\end{equation}
which is simply a statement of the K\"unneth theorem combined with the fact that twisted cohomology is supported on the middle dimension. 

\begin{figure}[H]
\centering
\includegraphics[width=0.5\textwidth]{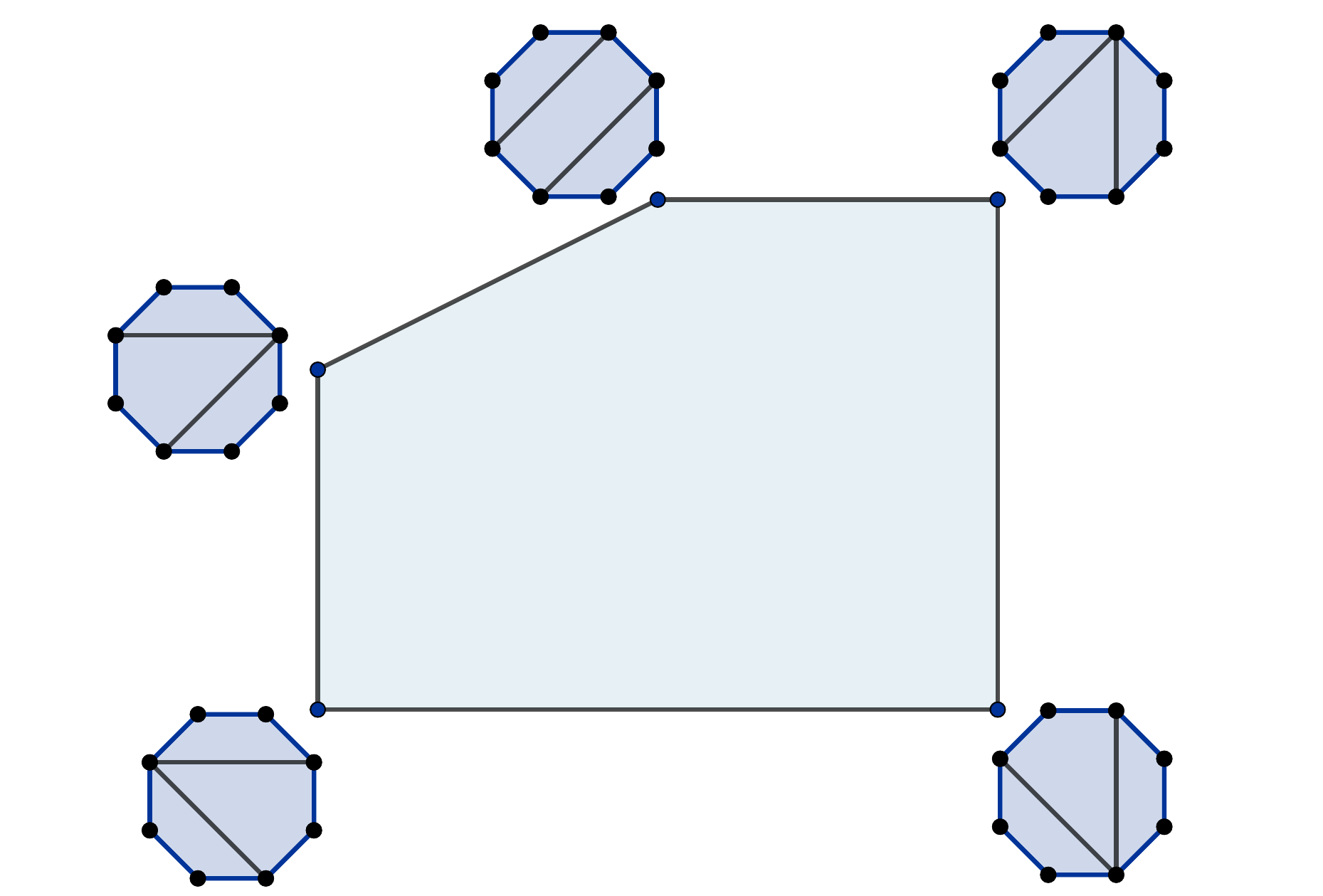}
\caption{Accordiohedron for Dissection $(14,47)$}\label{fig2.6}
\end{figure}

Now we deal with the second case - that of the dissection $(14,47)$. Here, the accordiohedron is a pentagon, with the vertices labelled by the quadrangulations $(14,47)$, $(38,47)$, $(36,38)$, $(16,36)$ and $(14,16)$. This is illustrated in figure \ref{fig2.6}.

The hyperplane arrangement furnishing this polytope in $\mathbb{CP}^{2}$ is,

\begin{equation}\label{eq2.14}
    \begin{aligned}
    f_{1} &= 1-x\\
    f_{2} &= 1+y\\
    f_{3} &= 2+x\\
    f_{4} &= 2+x-y\\
    f_{5} &= 2-y.
    \end{aligned}
\end{equation}
The twist capturing the correct kinematics is

\begin{equation}
    \omega_{(14,47)} = X_{36}d\ln f_{1} + X_{16}d\ln f_{2} + X_{14}d\ln f_{3} + X_{47}d\ln f_{4} + X_{38}d\ln f_{5},
\end{equation}
which can be verified by inspection of figure \ref{fig2.6}. We now encounter the first genuinely nontrivial scattering equations. They are as follows,

\begin{equation}
    \frac{X_{36}}{1-x} - \frac{X_{14}}{2+x} - \frac{X_{47}}{2+x-y} = 0
\end{equation}
and
\begin{equation}
    \frac{X_{36}}{1+y} - \frac{X_{47}}{2+x-y} - \frac{X_{38}}{2-y} = 0. 
\end{equation}

Our task now is to determine the number of solutions for these equations. As it turns out, the problem is actually computationally tedious. The \texttt{Solve} routine in {\scshape Mathematica} may be employed, and this determines the existence of four solutions for generic kinematics. A simpler method to do this however is to count the number of bounded chambers of the arrangement. This can be done far more readily, and the answer agrees with the former method. Thus, we note

\begin{equation}
    \mathrm{dim} H^{2}_{\omega_{(14,47)}}(\mathcal{A}_{(14,47)}) = 4.
\end{equation}
We note here that the \texttt{euler} routine in \texttt{Macaulay2} may be used to similar effect. This will prove useful when we consider three dimensional cases, where both of the previous methods - the counting of bounded chambers and brute force solving prove to be prohibitive.

Finally, before moving on to the next example, we point out that there are a total of $12$ quadrangulations of the $8$-gon. $4$ of these give rise to square type accordiohedra and $8$ give rise to pentagons.

\paragraph{$6$ Particles in $\phi^3$ + $\phi^4$}
The case of $6$-particle scattering in $\phi^3+\phi^4$ gives rise to a two dimensional accordiohedron when the process involves two trivalent and one quartic vertex. In this case, which was first studied in \cite{Jagadale:2019byr}, it was found that there were four primitive dissections, from which all of the others could be obtained by permutation. These are $(13,14)$, $(13,15)$, $(13,36)$ and $(13,46)$. 

The first three of these give rise to pentagons and the last one gives rise to a square. Since the combinatorics are essentially identical to that of the previous section, we simply list the accordiohedra in the images below.

\begin{figure}[H]
\centering
\begin{minipage}{.5\textwidth}
  \centering
  \includegraphics[width=0.8\textwidth]{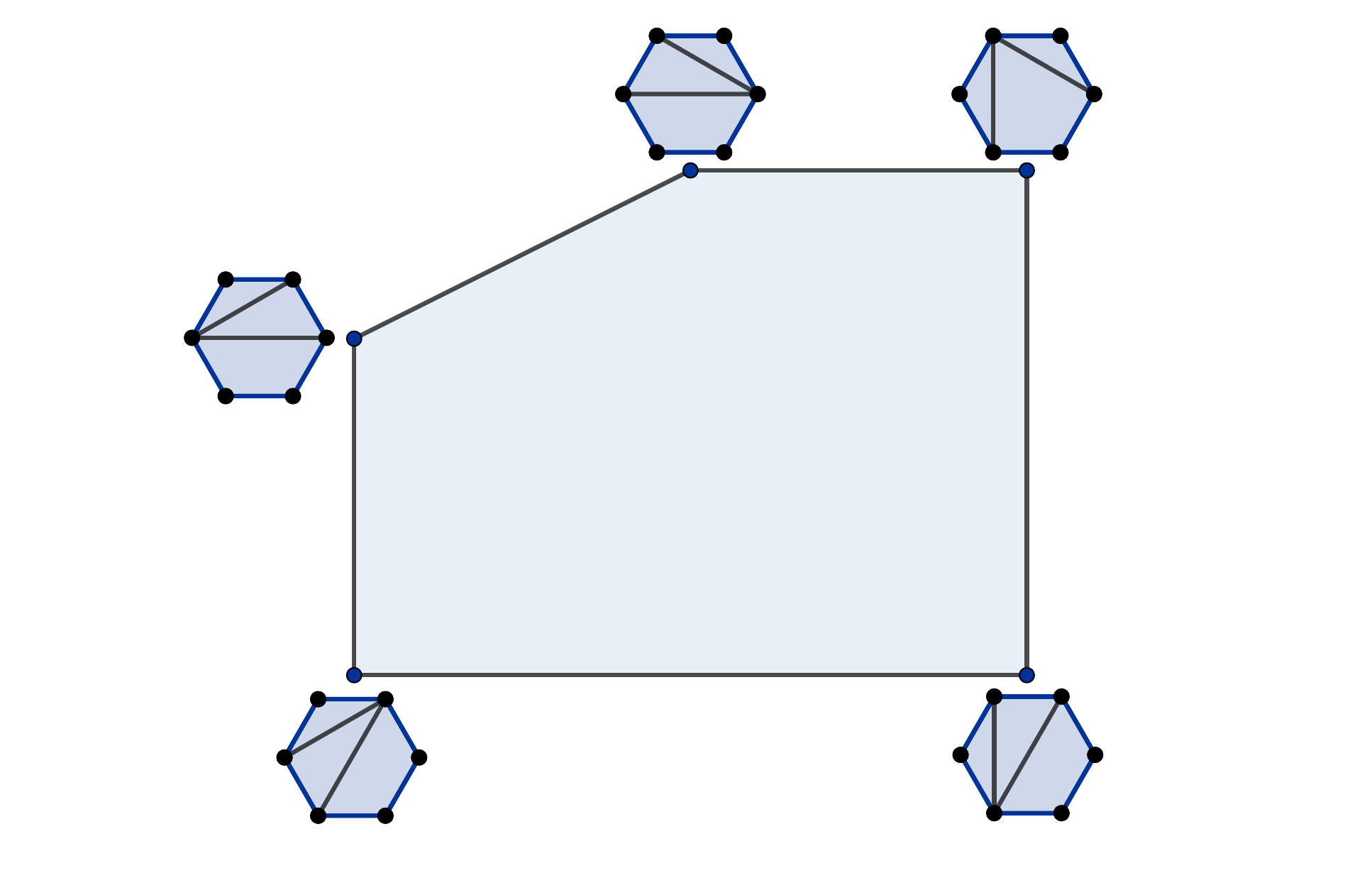}
  \captionof{figure}{Accordiohedron for Dissection $(13,14)$}
  \label{fig2.7}
\end{minipage}%
\begin{minipage}{.5\textwidth}
  \centering
  \includegraphics[width=0.8\linewidth]{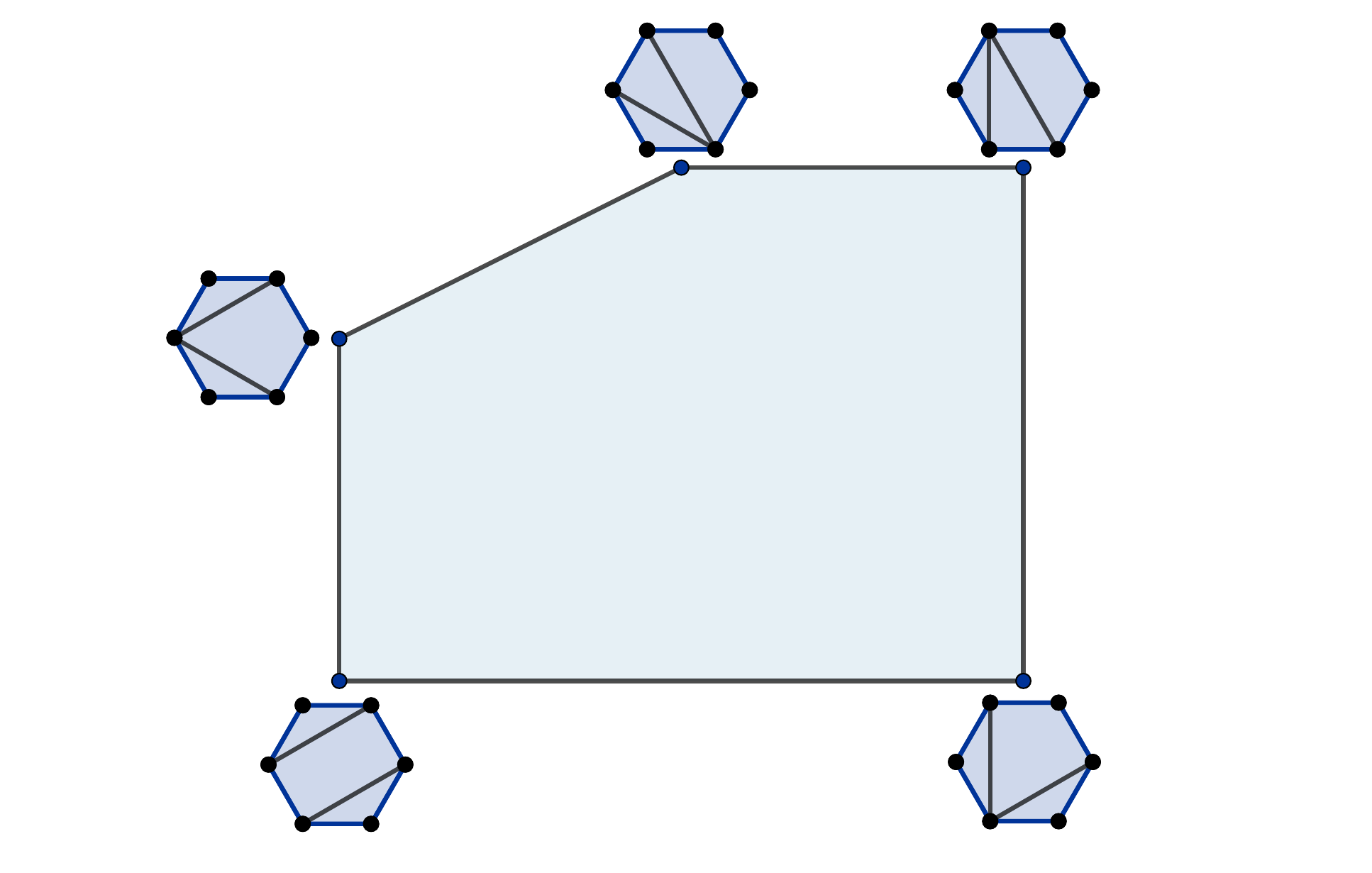}
  \captionof{figure}{Accordiohedron for Dissection $(13,15)$}
  \label{fig2.8}
\end{minipage}
\end{figure}

\begin{figure}[H]
\centering
\begin{minipage}{.5\textwidth}
  \centering
  \includegraphics[width=0.8\textwidth]{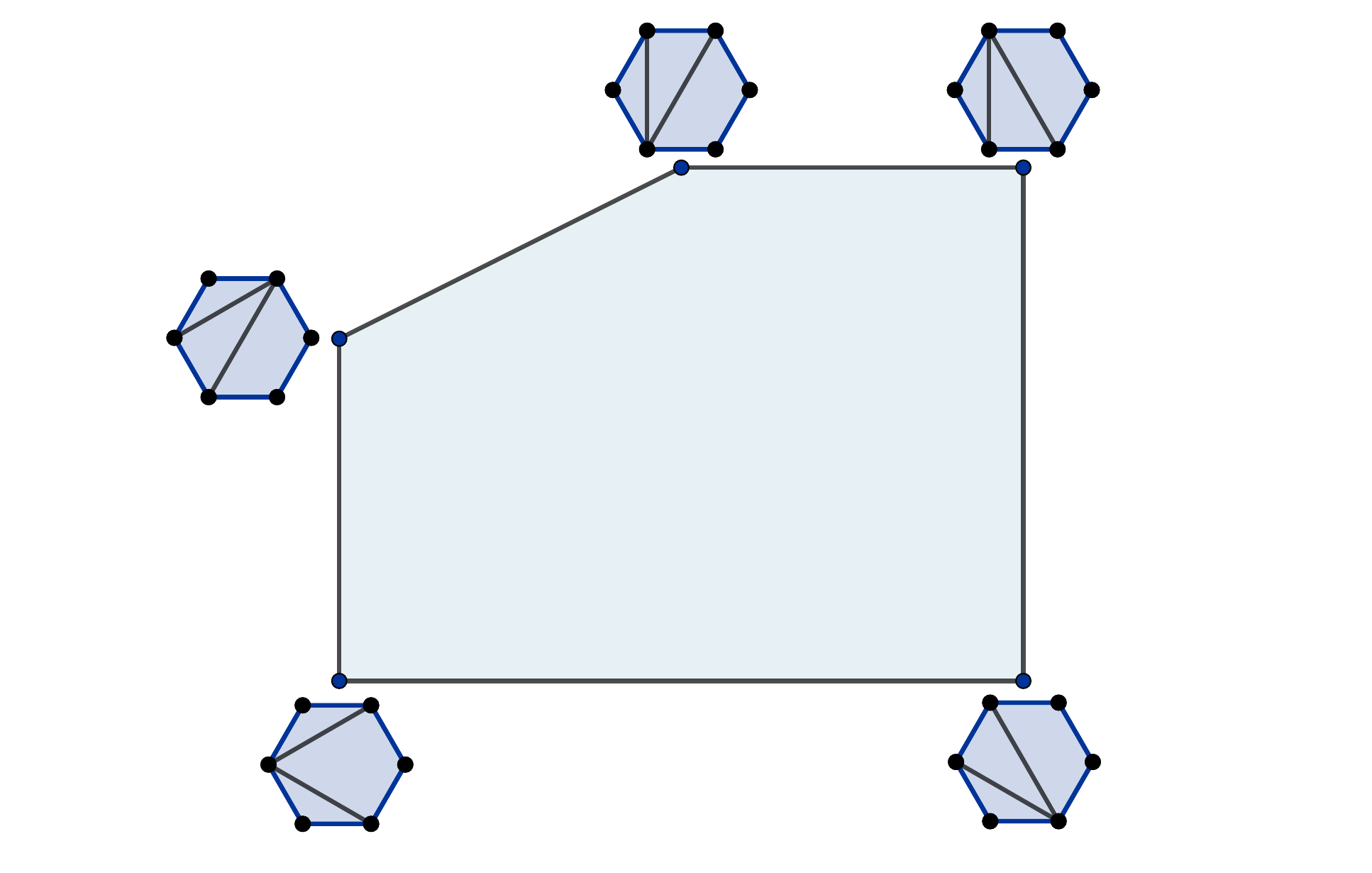}
  \captionof{figure}{Accordiohedron for Dissection $(13,36)$}
  \label{fig2.9}
\end{minipage}%
\begin{minipage}{.5\textwidth}
  \centering
  \includegraphics[width=0.8\linewidth]{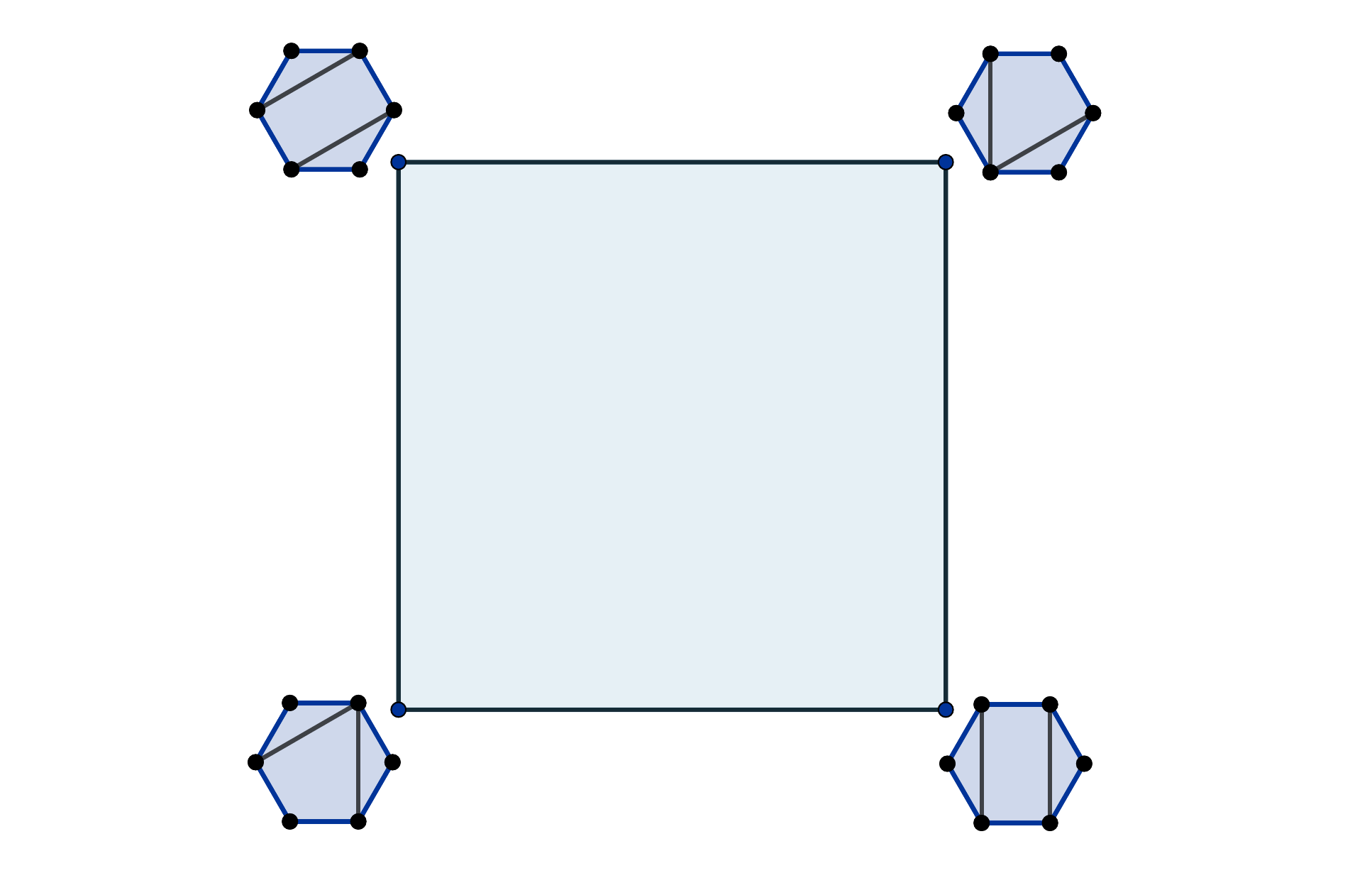}
  \captionof{figure}{Accordiohedron for Dissection $(13,46)$}
  \label{fig2.10}
\end{minipage}
\end{figure}

It should be clear from this that the hyperplane arrangements effecting these polytopes are identical to the arrangements considered in the previous section. For the polytopes in figures \ref{fig2.7}, \ref{fig2.8} and \ref{fig2.9}, we can use the arrangement in (\ref{eq2.14}) and for the polytope in figure \ref{fig2.10}, we use arrangement (\ref{eq2.8}). For the sake of completeness, let us list all the scattering equations involved. For figure \ref{fig2.7} we have,

\begin{equation}
    \begin{aligned}
    \omega_{(13,14),x}&:= \frac{X_{26}}{1-x} - \frac{X_{13}}{2+x} - \frac{X_{14}}{2+x-y} = 0\\
    \omega_{(13,14),y}&:=\frac{X_{35}}{1+y} - \frac{X_{14}}{2+x-y} - \frac{X_{24}}{2-y} = 0.
    \end{aligned}
\end{equation}
We have similarly for figure \ref{fig2.8},

\begin{equation}
    \begin{aligned}
    \omega_{(13,15),x}&:= \frac{X_{26}}{1-x} - \frac{X_{13}}{2+x} - \frac{X_{15}}{2+x-y} = 0\\
    \omega_{(13,15),y}&:=\frac{X_{46}}{1+y} - \frac{X_{15}}{2+x-y} - \frac{X_{24}}{2-y} = 0,
    \end{aligned}
\end{equation}
and for figure \ref{fig2.9}
\begin{equation}
    \begin{aligned}
    \omega_{(13,36),x}&:= \frac{X_{24}}{1-x} - \frac{X_{13}}{2+x} - \frac{X_{36}}{2+x-y} = 0\\
    \omega_{(13,36),y}&:=\frac{X_{15}}{1+y} - \frac{X_{36}}{2+x-y} - \frac{X_{26}}{2-y} = 0.
    \end{aligned}
\end{equation}
Finally for figure \ref{fig2.10} we have,

\begin{equation}
    \begin{aligned}
    \omega_{(13,46),x}&:= \frac{X_{26}}{1-x} - \frac{X_{13}}{2+x}=0\\
    \omega_{(13,46),y}&:=\frac{X_{35}}{1+y}  - \frac{X_{46}}{2-y} = 0.
    \end{aligned}
\end{equation}
From these we infer the dimensions

\begin{equation}
    \mathrm{dim}H_{\omega_{(13,14)}}^{2}(\mathcal{A}_{(13,14)}) = \mathrm{dim}H_{\omega_{(13,15)}}^{2}(\mathcal{A}_{(13,15)}) = \mathrm{dim}H_{\omega_{(13,36)}}^{2}(\mathcal{A}_{(13,36)}) = 4
\end{equation}
and
\begin{equation}
    \mathrm{dim}H_{\omega_{(13,46)}}^{2}(\mathcal{A}_{(13,46)}) = 1.
\end{equation}

\subsection{Three Dimensional Polytopes}
\paragraph{$10$ Particles in $\phi^4$}
The last case we consider is the scattering of $10$-particles with quartic interactions. This case is far more nontrivial than those considered till this point. For the first time, we encounter three-dimensional accordiohedra, and the counting of the number of solutions of the scattering equations becomes more involved. This case was originally considered in \cite{Banerjee:2018tun} when Stokes polytopes were applied for the first time to scattering amplitudes. In this section, we review the combinatorial details and provide precise convex realizations of these polytopes and the relevant scattering equations. In this section, we will omit the onerous details involving the vertices of the accordiohedra and only present data about the facets. 

\begin{figure}[H]
\centering
\includegraphics[width=0.4\textwidth]{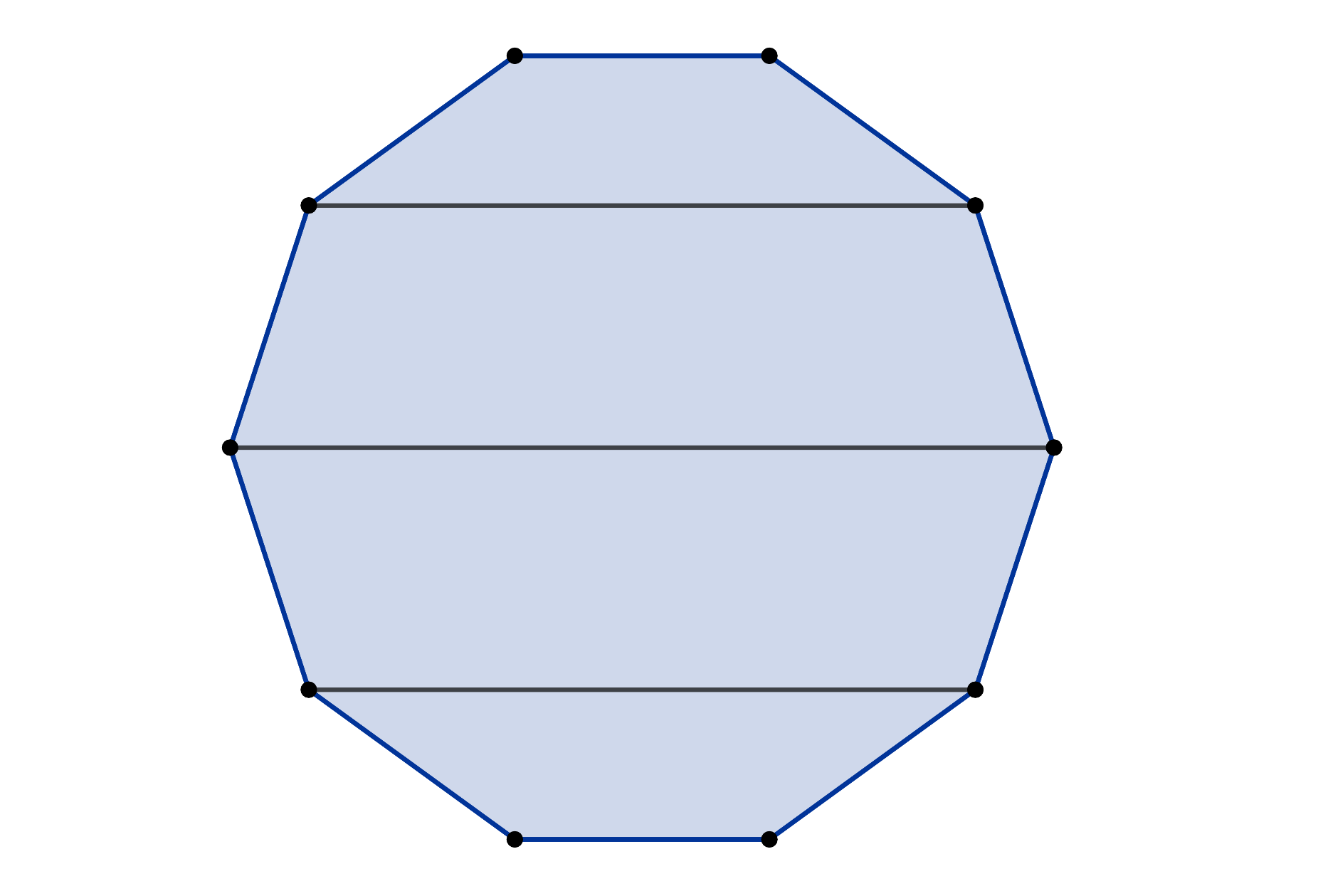}
\caption{Dissection $(14,510,69)$}\label{fig2.11}
\end{figure}

For this scattering process, there are four classes of accordiohedra that appear. The simplest, known as cube type, arises out of the quadrangulation $(14,510,69)$, which has been illustrated in figure \ref{fig2.11}.

\begin{figure}[H]
\centering
\includegraphics[width=0.29\textwidth]{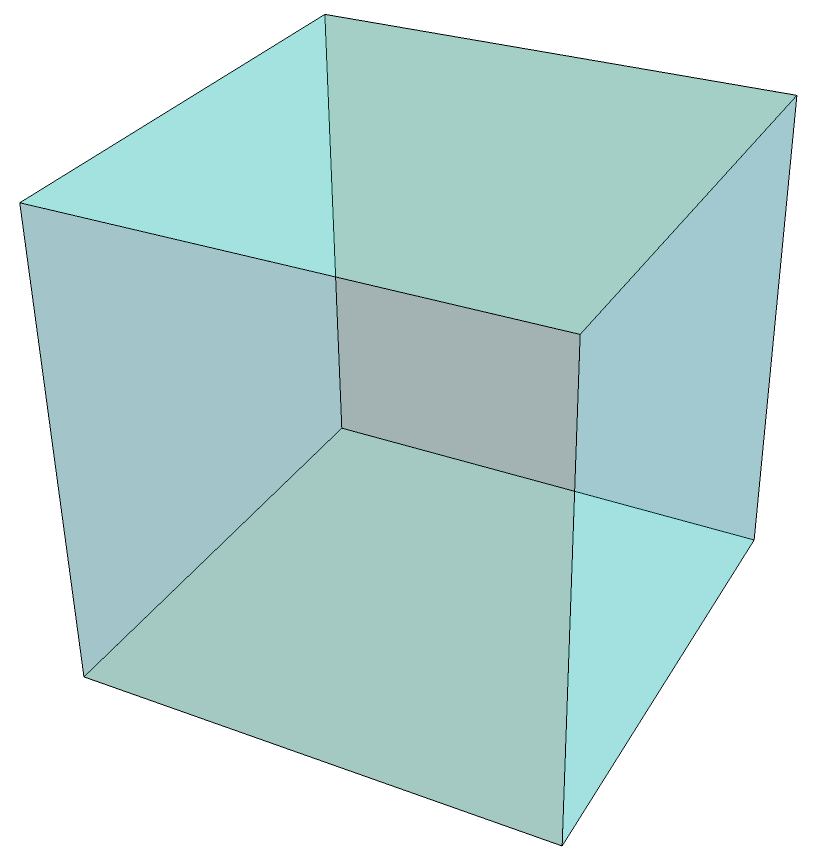}
\caption{Accordiohedron for Dissection $(14,510,69)$}\label{fig2.12}
\end{figure}

The accordiohedron for this dissection has six facets, labelled by partial quadrangulations $(14)$, $(510)$, $(69)$, $(310)$, $(49)$ and $(58)$, realized as a convex polytopes by the following hyperplanes $f_{1}$ through $f_{6}$ in $\mathbb{CP}^{3}$

\begin{equation}
    \begin{aligned}
    &f_{1} = x - 3  \hspace{1cm} &f_{2} = y - 3 \\
    &f_{3} = z - 3 \hspace{1cm}  &f_{4} = x + 2 \\ 
    &f_{5} = y + 2 \hspace{1cm} &f_{6} = z +2. 
   \end{aligned}
\end{equation}
This arrangement is presented in figure \ref{fig2.12}.

The scattering equations are now just three copies of the scattering equation for a line,

\begin{equation}
    \omega_{(14,510,69),x} := \frac{X_{14}}{x-3} - \frac{X_{310}}{x+2} = 0,
\end{equation}

\begin{equation}
    \omega_{(14,510,69),y} := \frac{X_{510}}{y-3} - \frac{X_{49}}{y+2} = 0
\end{equation}
and

\begin{equation}
    \omega_{(14,510,69),z} := \frac{X_{510}}{z-3} - \frac{X_{58}}{z+2} = 0.
\end{equation}
The K\"unneth formula may be readily applied to see (trivially so) that,

\begin{equation}
    \mathrm{dim}H^{3}_{\omega_{(14,510,69)}}(\mathcal{A}_{(14,510,69)}) = 1
\end{equation}
corresponding to one solution of the scattering equations.

\begin{figure}[H]
\centering
\includegraphics[width=0.4\textwidth]{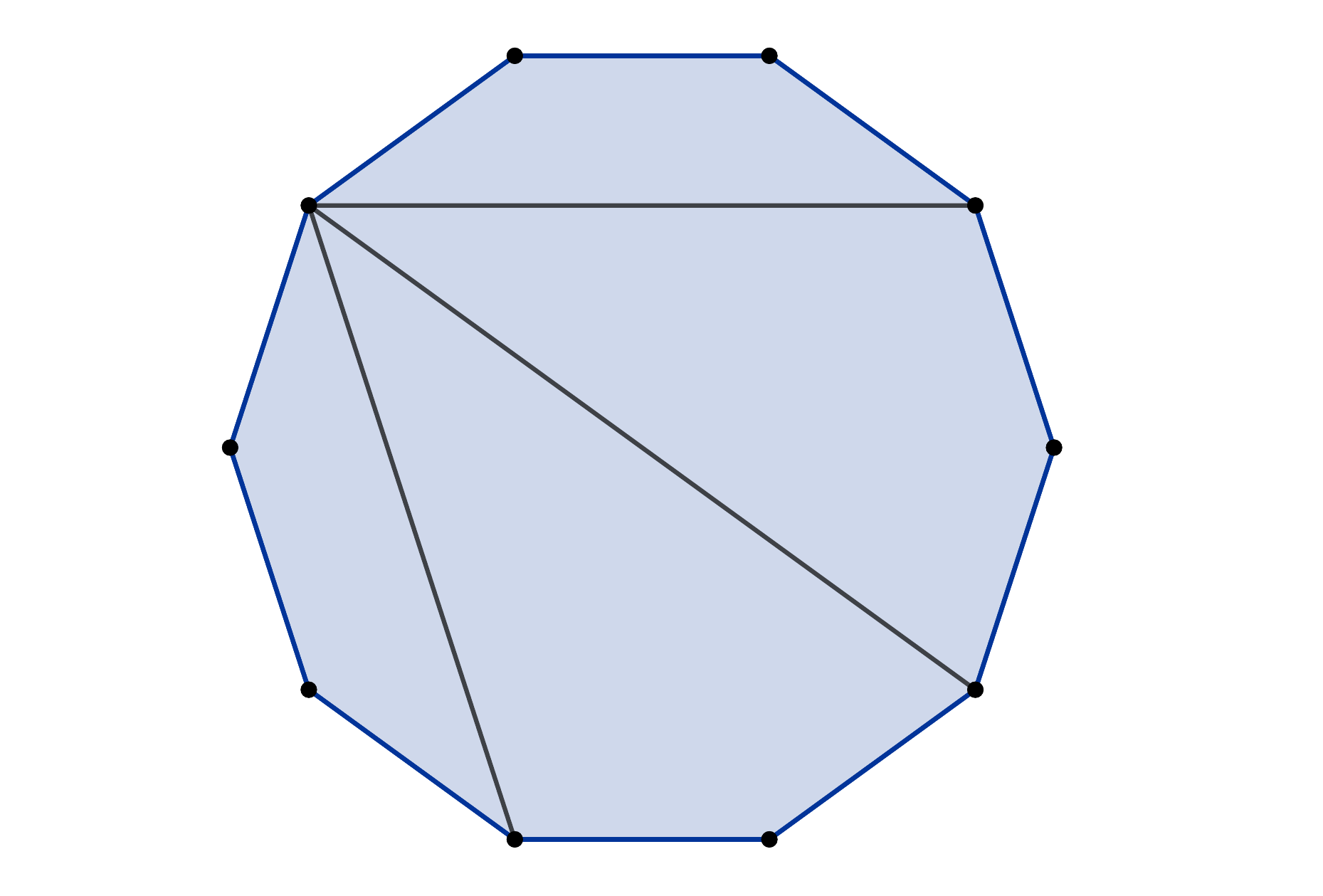}
\caption{Dissection $(14,16,18)$}\label{fig2.13}
\end{figure}

The second type of accordiohedron is a three dimensional associahedron, arising from three types of primitive quadrangulations, namely $(14,16,18)$, $(14,16,69)$ and $(14,49,69)$. The latter two are entirely analogous to the first case. Accordingly, we focus on the former in this discussion. Diagramatically we have the quadrangulation shown in \ref{fig2.13}.

\begin{figure}[H]
\centering
\includegraphics[width=0.4\textwidth]{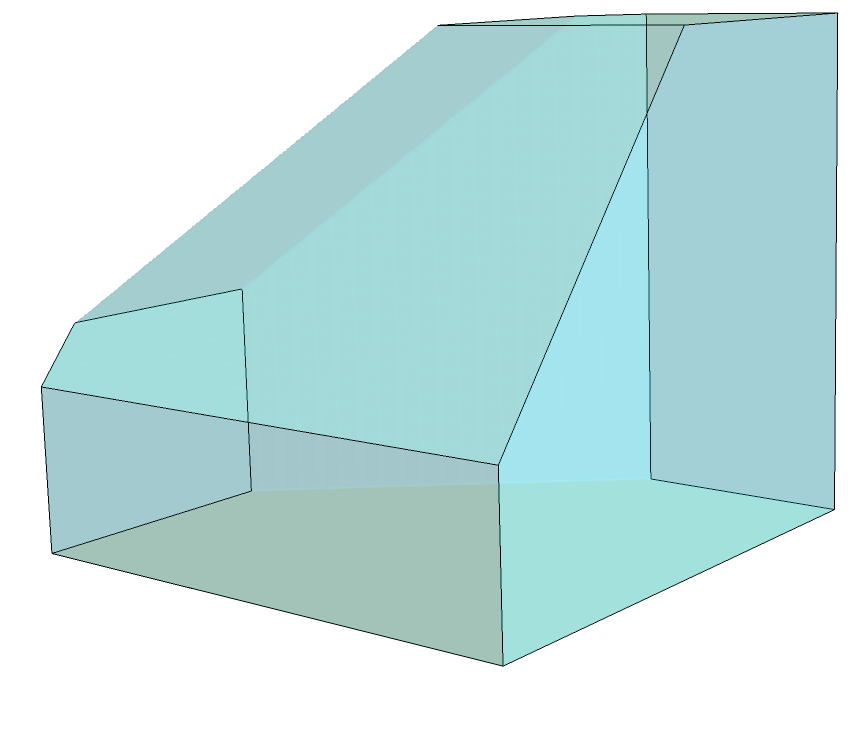}
\caption{Accordiohedron for Dissection $(14,16,18)$}\label{fig2.14}
\end{figure}

The facets of the accordiohedron generated by this dissection are $9$ in number, and are labelled by the partial dissections $\lbrace{14,16,18,36,58,710,38,510,310\rbrace}$. The corresponding hyperplane arrangement $\lbrace{f_{1},...,f_{9}\rbrace}$ is as follows,

\begin{equation}
    \begin{split}
    &f_{1} = x - 4  \\
    &f_{3} = z-6\\
    &f_{5} = z-y-6\\ 
    &f_{7} = z-x-7 \\
    &f_{9} = x+6\\
   \end{split}\qquad
   \begin{split}
       &f_{2} = y - 6 \\
    &f_{4} = y - x - 6 \\ 
     &f_{6} = z+3 \\
    &f_{8} = y+6 \\
   \end{split}
\end{equation}
which yields figure \ref{fig2.14}.

The scattering equations for this arrangement are given as follows

\begin{equation}
    \begin{aligned}
    \omega_{(14,16,18),x} := \frac{X_{14}}{x-4} -\frac{X_{310}}{x+6} -\frac{X_{36}}{x+6-y} -\frac{X_{38}}{x-z+7} = 0,
    \end{aligned}
\end{equation}

\begin{equation}
    \begin{aligned}
    \omega_{(14,16,18),y} := \frac{X_{16}}{y-5}- \frac{X_{510}}{y+6} +\frac{X_{36}}{y-6+x} -\frac{X_{58}}{y-z+6} = 0
    \end{aligned}
\end{equation}
and

\begin{equation}
    \begin{aligned}
    \omega_{(14,16,18),z} := \frac{X_{18}}{z-6} -\frac{X_{710}}{z+3} + \frac{X_{58}}{z-y-6} + \frac{X_{38}}{z-x-7} = 0.
    \end{aligned}
\end{equation}
In order to find the number of solutions of this system of equations, two approaches may be employed. One possible method is direct computation on \texttt{Solve} in {\scshape Mathematica}. This method however was prohibitive due to the CPU time required. A simpler approach involves using the \texttt{euler} routine in the \texttt{HyperplaneArrangements} package in \texttt{Macaulay2}. The Euler characteristic of this hyperplane arrangement computed in this manner is $24$. We thus have,

\begin{equation}
    \mathrm{dim}H^{3}_{\omega_{(14,16,18)}}\left(\mathcal{A}_{(14,16,18)}\right) = 24.
\end{equation}
This implies that there are $24$ bounded chambers for this hyperplane arrangement.

We consider next the case of the dissection $(14,47,710)$, which gives rise to an accordiohedron of so-called Lucas type, which is a genuine accordiohedron in the sense that it cannot be understood as an associahedron. 

\begin{figure}[H]
\centering
\includegraphics[width=0.4\textwidth]{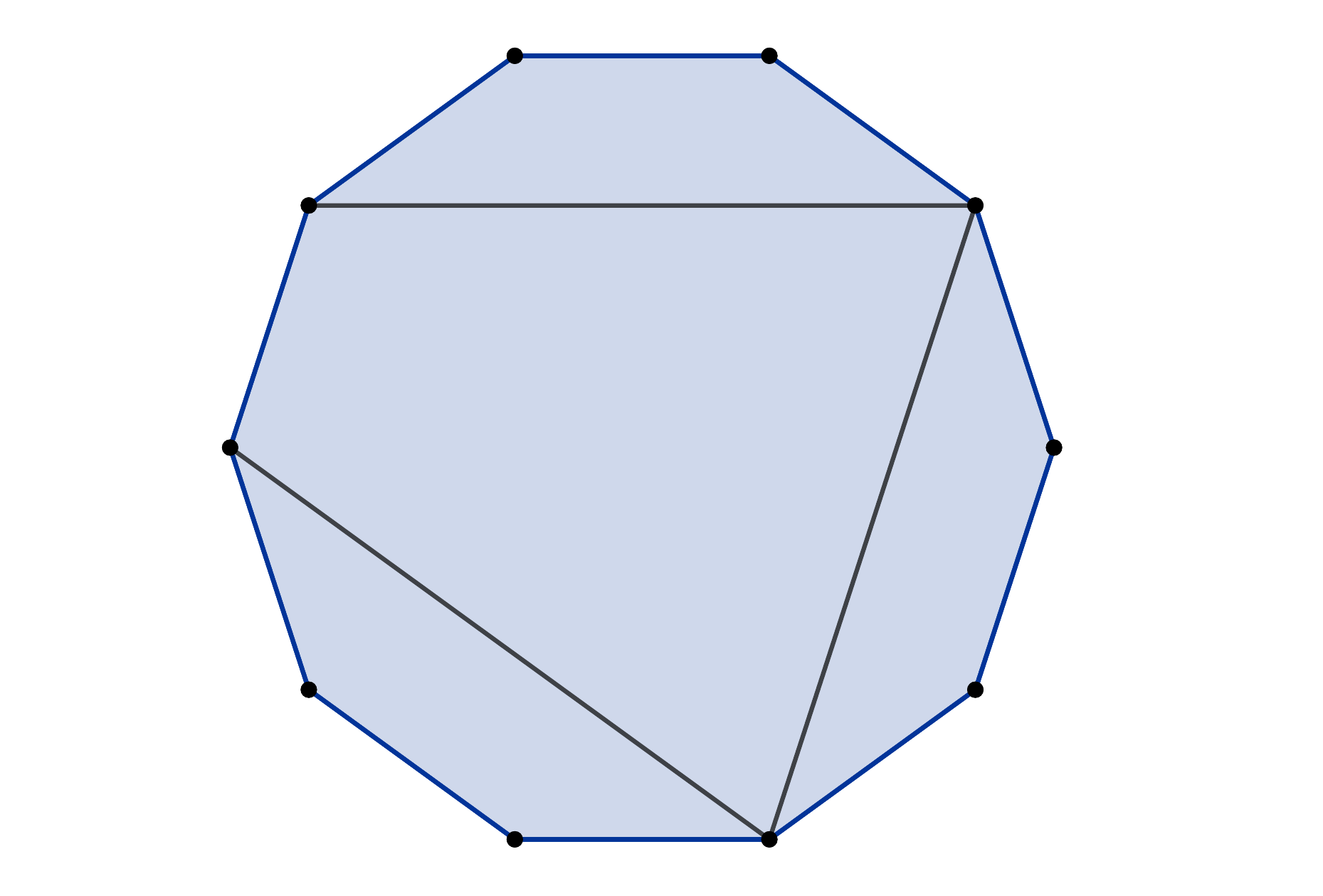}
\caption{Dissection $(14,47,710)$}\label{fig2.15}
\end{figure}

\begin{figure}[H]
\centering
\includegraphics[width=0.4\textwidth]{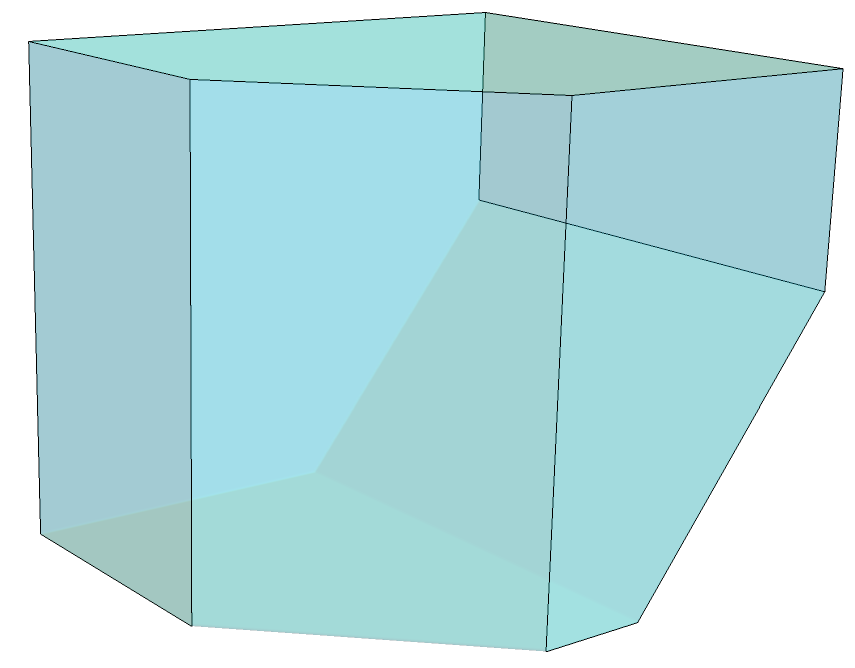}
\caption{Accordiohedron for Dissection $(14,47,710)$}\label{fig2.16}
\end{figure}

This polytope, obtained by finding all dissections of a $10$-gon with three quadrangulations compatible with $(14,47,710)$ has $12$ vertices, with $8$ facets, labelled by $ \lbrace{14,47,710,16,49,310,36,69\rbrace}$. The corresponding hyperplane arrangement in $\mathbb{CP}^{3}$ is

\begin{equation}
    \begin{split}
    &f_{1} = x - 5  \\
    &f_{3} = y-4\\
    &f_{5} = y-z-5\\ 
    &f_{7} = y+6 \\
   \end{split}\qquad
   \begin{split}
       &f_{2} = y - 5 \\
    &f_{4} = x-y-6 \\ 
     &f_{6} = x+4 \\
    &f_{8} = z+4. \\
   \end{split}
\end{equation}
Visually this is represented in figure \ref{fig2.16}.

The scattering equations can be inferred from the twist, which is determined by the kinematic assignements. They are

\begin{equation}
    \omega_{(14,47,710),x}:= \frac{X_{14}}{x-5} -\frac{X_{16}}{x-y-6} -\frac{X_{310}}{x+4} = 0,
\end{equation}

\begin{equation}
    \omega_{(14,47,710),y}:= \frac{X_{47}}{y-5} +\frac{X_{16}}{y-x+6} -\frac{X_{36}}{x+4} +\frac{X_{49}}{y-z-5} = 0, 
\end{equation}
and

\begin{equation}
    \omega_{(14,47,710),z}:= \frac{X_{710}}{z-4} -\frac{X_{49}}{z-y+5} -\frac{X_{69}}{z+4} = 0. 
\end{equation}
The number of solutions to this system was determined in \texttt{Macaulay2} to be $12$. Hence we have,

\begin{equation}
    \mathrm{dim}H^{3}_{\omega_{(14,47,710)}}\left(\mathcal{A}_{(14,47,710)}\right) = 12
\end{equation}
following which we may state that there are $12$ bounded chambers in this arrangement. 

The final case we have to consider is known as the mixed accordiohedron, which is generated by the quadrangulations $(14,510,710)$ and $(14,16,17)$. Again, we focus on the former case for convenience. 

\begin{figure}[H]
\centering
\includegraphics[width=0.4\textwidth]{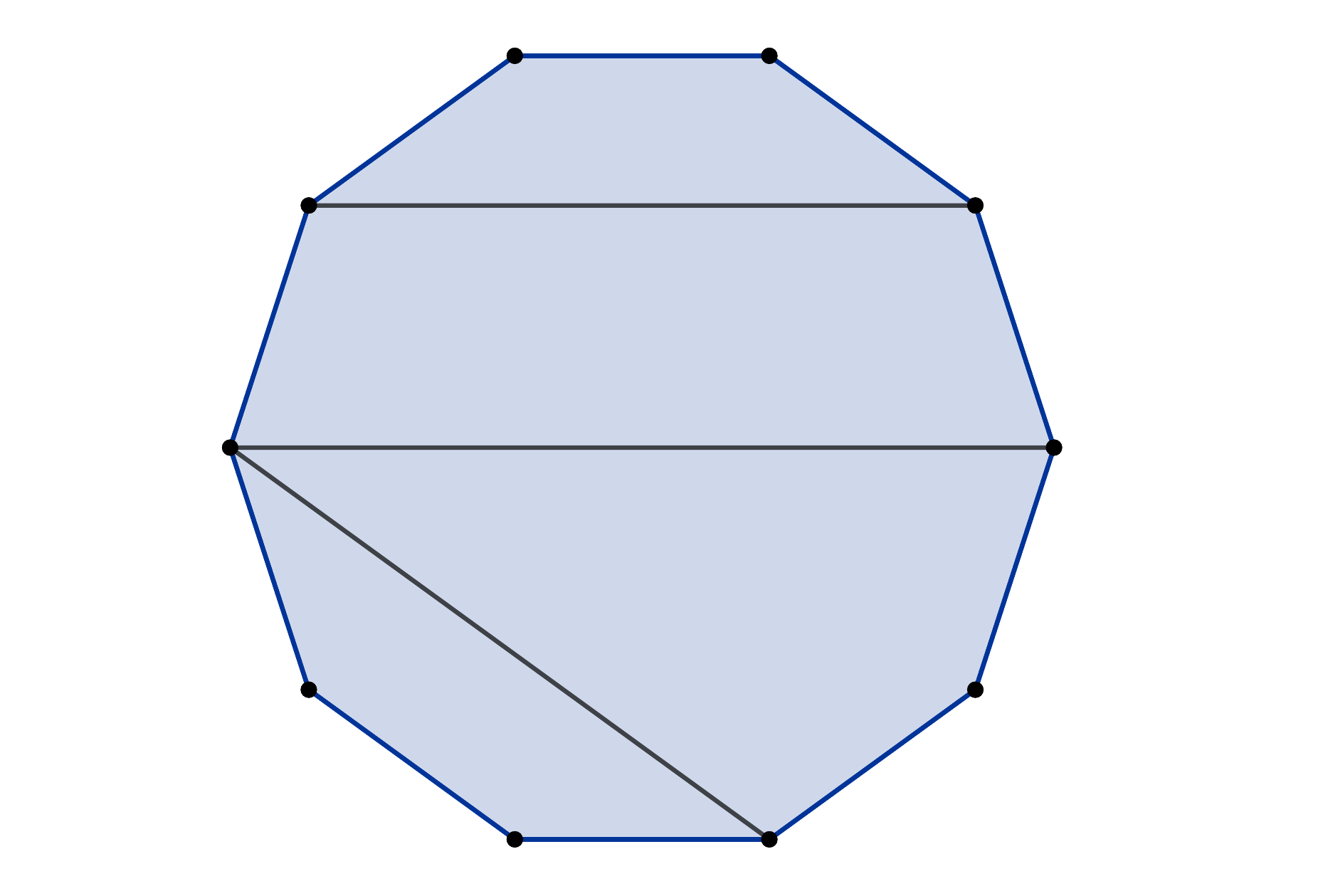}
\caption{Dissection $(14,510,710)$}\label{fig2.17}
\end{figure}

\begin{figure}[H]
\centering
\includegraphics[width=0.4\textwidth]{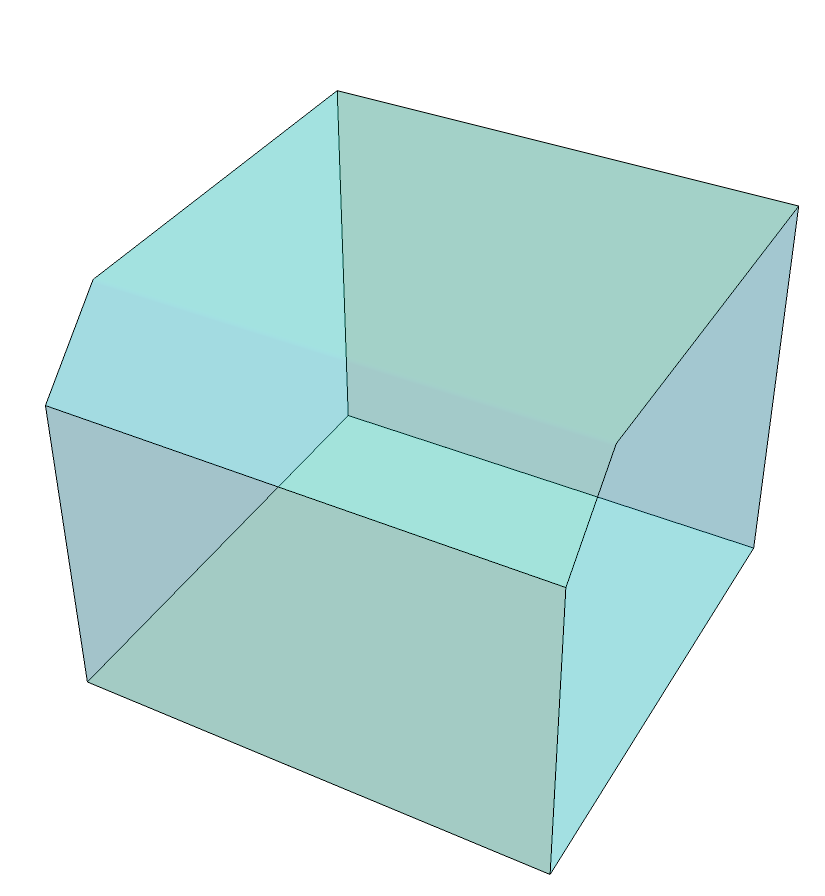}
\caption{Accordiohedron from Dissection $(14,510,710)$}\label{fig2.18}
\end{figure}

The accordiohedron in this case actually ends up being a product polytope, formed out of a pentagon times a line. Accordingly, we have $7$ facets, labelled by the dissections $\lbrace{14,510,710,310,57,69,49\rbrace}$. This convex polytope is afforded by the arrangement

\begin{equation}
        \begin{split}
    &f_{1} = x - 4  \\
    &f_{3} = z-2\\
    &f_{5} = z-y-4\\ 
    &f_{7} = y+4 \\
   \end{split}\qquad
   \begin{split}
       &f_{2} = y - 3 \\
    &f_{4} = x+3 \\ 
     &f_{6} = z+4 \\
   \end{split}
\end{equation}
which produces the polytope shown in figure \ref{fig2.18}.

One can observe from the diagram that this is indeed a pentagon times a line. Accordingly, the K\"unneth formula is applied readily to obtain,

\begin{equation}
    \mathrm{dim}H^{3}_{\omega_{(14,510,710)}}\left(\mathcal{A}_{(14,510,710)}\right) = 4\times 1 = 4.
\end{equation}
For the sake of completeness, we include the scattering equations,

\begin{equation}
    \omega_{(14,510,710),x}:= \frac{X_{14}}{x-4} - \frac{X_{310}}{x+3} = 0, 
\end{equation}

\begin{equation}
    \omega_{(14,510,710),y}:= \frac{X_{510}}{y-3} - \frac{X_{49}}{y+4} - \frac{X_{57}}{y-z+4} = 0,
\end{equation}
and

\begin{equation}
    \omega_{(14,510,710),z}:= \frac{X_{710}}{z-2} - \frac{X_{69}}{z+4} + \frac{X_{57}}{z-y-4} = 0. 
\end{equation}
These cases exhaust all the accordiohedra that arise for this scattering process, as all of the others can be obtained by permuting the indices of the primitives considered. 

Let us summarize the salient results of this section before moving on. What we have attended to so far is a class of convex realizations for a number of accordiohedra considered in the literature in one, two and three dimensions. Rather than treating them in the sense of the traditional positive geometry program, we have examined these constructions in the general context of twisted intersection theory. While we have not yet evaluated any amplitudes explicitly, we have provided the data needed to do so, namely the hyperplane arrangements supplying the polytopes, as well as the twist differential forms that encode the kinematical structure of the scattering processes. The components of the twist have been interpreted as the analogue of the scattering equations of Cachazo, He and Yuan for the more complicated interactions that we want to study. 

\pagebreak

\section{The KLT Relations for Accordiohedra}\label{sec:relations}

The double copy relations due to Kawai, Lewellen and Tye were first developed in the study of string amplitudes. They observed that the scattering of $n$ massless closed string states could be rearranged as a product of two open string scattering amplitudes of $n$ massless states each. This product however cannot be carried out blindly. An object known as the KLT kernel must be used to convolve $(n-3)!$ open string amplitudes with another copy of $(n-3)!$ amplitudes, where the permutations are carried out over two choices of $n-3$ particles for each of the open string amplitudes.

The original derivation of the KLT relations used contour deformation arguments, which combined with the splitting into holomorphic and antiholomorphic parts of the closed string integrand breaks the integral over $\mathcal{M}_{0,n}$ into two integrals over the $n$-simplex $\Delta_{n}$. The trade-off is the introduction of the KLT kernel, which fuses the two open string integrals together. Schematically we have,

\begin{equation}
    \int_{\mathcal{M}_{0,n}} |\mathcal{I}|^{2} = \sum_{\alpha,\beta}\left(\int_{\Delta_{n}}\mathcal{I}(\alpha)\right)K[\alpha|\beta]\left(\int_{\Delta_{n}}\mathcal{I}(\beta)\right). 
\end{equation}
Here, the closed string integrand has been factorized into two open string integrals. Doing this requires the introduction of two bases $\alpha$ and $\beta$, which run over two arbitrary permutations of the $n$ particles, playing the role of a colour ordering. In the field theory limit, this expression reduces to a relationship between the scattering of gravitons and Yang-Mills amplitudes at tree level. We will focus on this avatar of the KLT kernel henceforth.

While the KLT relation in field theory can be derived in a number of ways, the approach that we are most interested in is that of twisted intersection theory. To review the basic details involved, consider the moduli space of marked Riemann spheres $\mathcal{M}_{0,n}$. Let us equip this space with the following differential form,

\begin{equation}
    \omega = \sum_{i<j}s_{ij}d\ln(z_{i}-z_{j}),
\end{equation}
where $s_{ij} = (p_i+p_j)^2$, and all particles are considered massless. This form acts as a \emph{twist}, which modifies the exterior derivative to

\begin{equation}
    \nabla_{\omega} = d + \omega\wedge .
\end{equation}
This twisted differential operator can be verified to be nilpotent, which leads to a natural generalization of the de-Rham complex into the twisted de Rham complex and a corresponding cohomology theory, which can be defined by,

\begin{equation}\label{eq3.4}
    H^{k}\left(\mathcal{M}_{0,n},\nabla_{\omega}\right) = \frac{\lbrace{h\in \Omega^{k}\left(\mathcal{M}_{0,n}\right)|\nabla_{\omega}h = 0\rbrace}}{\lbrace{h\in \Omega^{k}\left(\mathcal{M}_{0,n}\right)|h = \nabla_{\omega }f\rbrace}}. 
\end{equation}
An important theorem due to Aomoto established that this cohomology theory is trivial for all $k\neq (n-3)$\footnote{Note that we have $n \geq 3$.}, which is middle dimensional. 

A basis of this cohomology class which is particularly valuable for the study of scattering amplitudes is known as the Parke-Taylor basis. For a given permutation $\sigma\in S_{n-3}$ the following form may be defined,

\begin{equation}
    \mathrm{PT}(\sigma) = \frac{dz_1\wedge ... \wedge dz_{n}}{(z_{1}-z_{2})(z_{2}-z_{\sigma(3)})(z_{\sigma(3)}-z_{\sigma{4}})...(z_{n}-z_{\sigma(n-1)})(z_{1}-z_{1})}.
\end{equation}
It can be shown (a key result of \cite{Mizera:2017cqs}) that this collection of $(n-3)!$ forms furnishes a complete basis for $H^{(n-3)}\left(\mathcal{M}_{0,n},\nabla_{\omega}\right)$. 
Given two representatives $\varphi_1$ and $\varphi_2$ of the cohomology group in (\ref{eq3.4}), it was observed in a number of studies \cite{cho1995,Mimachi_2003,yoshida1,yoshida2,yoshida3,AOMOTO1997119,AOMOTO-KITA,YOSHIDA} that a pairing $\langle{\varphi_1,\varphi_2\rangle}$ may be defined between the two, which acts as a twisted generalization of the traditional notion of an intersection number. These ideas found their first application to field theory when Mizera proved in \cite{Mizera:2017rqa} that this pairing between two Parke-Taylor forms $\mathrm{PT(\alpha)}$ and $\mathrm{PT(\beta)}$ gives precisely the tree level planar scattering amplitude for the biadjoint scalar theory:

\begin{equation}
    \langle{\mathrm{PT}(\alpha),\mathrm{PT}(\beta)\rangle} = m(\alpha|\beta).
\end{equation}
To relate this fact to the KLT relations, one simply notes that the Pfaffian $\mathrm{Pf}'\Psi$, which arises in the CHY formalism can be recast in a Parke-Taylor basis. The pairing between a Pfaffian and Parke-Taylor factor then becomes precisely the CHY formula for Yang-Mills amplitudes, while the pairing between two Pfaffians becomes the CHY formula for gravity amplitudes. An application of the twisted period relations (\ref{eq1.3}) with $\varphi_1 = \varphi_2 = \mathrm{Pf}'\Psi$ with a Parke-Taylor basis for the cohomology group yields the KLT relations.

Our main goal in this section will be to repeat this analysis when we have hyperplane arrangements in $\mathbb{CP}^{n}$ that produce accordiohohedra rather than $\mathcal{M}_{0,n}$. In order to do this, we have to use the formalism developed in \cite{Kalyanapuram:2019nnf} and \cite{Kalyanapuram:2020tsr}. We will now recall some of the principal aspects of the techniques developed therein.

We start with a collection of hyperplanes, say $f_{1},...,f_{N}$ in $\mathbb{CP}^{n}$, which realizes an $n$-dimensional polytope $\mathcal{A}_{n}$\footnote{We will also use this symbol to denote the complement of the hyperplane arrangement; there should be no confusion on this point} as the interior of the arrangement. Given this, we can now define a twist $\omega_{\A_n}$, encoding kinematical data. The main result of the recent work of the author was that the form

\begin{equation}
    \varphi_{\mathcal{A}_{n}} = \sum_{f_{i_1}\cap f_{i_2}\cap...\cap f_{i_n}\in \mathrm{Vertex}}d\ln f_{i_{1}}\wedge ... \wedge d\ln f_{i_{n}}
\end{equation}
when viewed as an element of $H^{n}_{\omega_{\mathcal{A}_{n}}}\left(\mathcal{A}_{n},\nabla_{\omega_{\A_n}}\right)$ gives rise to the pairing $\langle{\varphi_{\A_n},\varphi_{\A_n}\rangle}$. The pairing was then shown to equal the scattering amplitude defined by the accordiohedron using the localization formula proved in \cite{Mizera:2017rqa}.

In this section, our aim will be to use this fact to develop a suitable generalization of the KLT relations for amplitudes derived from accordiohedra. Our basic technique will mirror the one used for associahedra, namely that we will find a basis for the twisted cohomology group and employ the twisted period relations. There are a couple of qualitative differences here that are worth noting however. An important fact that differentiates accordiohedra from associahedra is that we can have several accordiohedra for a given process, while associahedra are unique. Accordingly, we will obtain a family of KLT relations when working with these polytopes. Additionally, it seems that the elements of the KLT matrix in the case of accordiohedra are generically nonplanar amplitudes. There does not seem to be any canonical choice like the Parke-Taylor basis in the case of accordiohedra. Thus, for the time being, we are left with having to do brute force calculations. Clarifying the implications of these observations should be fertile ground for future research.

\subsection{One-Dimensional KLT Relations}
We start with the simplest example of KLT relations for accordiohedra - the case of one dimensional accordiohedra, which are basically associahedra. Since the essential aspects of these relations have already been worked out in \cite{Mizera:2017cqs}, the main point of this section will be to provide the correct analogue for those processes which admit geometric representations in terms of intersection numbers of one dimensional accordiohedra.

A one dimensional accordiohedron is simply a line, so we have the following hyperplane arrangement,

\begin{equation}
    \begin{aligned}
    f_{1} &= 0 \\
    f_{2} &= 1
    \end{aligned}
\end{equation}
in $\mathbb{CP}^{1}$. Thus, we are led to working on the complement

\begin{equation}
    \mathcal{A}_{1} = \mathbb{CP}^{1} - \lbrace{0,1,\infty\rbrace}.
\end{equation}
On this space, we now define a twist, which encodes kinematical information. To keep the discussion as general as possible, we will use the following twist,

\begin{equation}
    \omega_{\mathcal{A}_{1}} = \alpha_{1} d\ln(x) + \alpha_{2}d\ln(x-1).
\end{equation}
We have now the form,

\begin{equation}\label{eq3.12}
    \varphi_{\mathcal{A}_{1}} = d\ln\left(\frac{x}{x-1}\right)
\end{equation}
and the intersection number

\begin{equation}
    \langle{ \varphi_{\mathcal{A}_{1}}, \varphi_{\mathcal{A}_{1}}\rangle} = \frac{1}{\alpha_1} + \frac{1}{\alpha_2}.
\end{equation}
To find the KLT relations, we first note that the dimension is

\begin{equation}
    \mathrm{dim} H^{1}_{\omega_{\A_1}}\left(\A_1, \nabla_{\omega_{\A_1}}\right) = 1.
\end{equation}
Accordingly, use of the twisted Riemann period relations, which requires inserting two copies of bases, can be made by specifying a two basis elements, each of which separately furnishes a basis of the twisted cohomology group. We will look at three such examples.

\paragraph{Example 3.1}
Consider a basis for the twisted cohomology group given by the element $\phi_1 = \varphi_{\mathcal{A}_{1}}$. In this case, the KLT matrix becomes,

\begin{equation}
    \langle{\phi_1,\phi_1\rangle}^{-1} = \frac{\alpha_1\alpha_2}{\alpha_1+\alpha_2}.
\end{equation}
Applying the twisted Riemann period relations is now a trivial multiplication,

\begin{equation}
    \left(\frac{1}{\alpha_1} + \frac{1}{\alpha_2}\right)\frac{\alpha_1\alpha_2}{\alpha_1+\alpha_2}\left(\frac{1}{\alpha_1} + \frac{1}{\alpha_2}\right) = \frac{1}{\alpha_1} + \frac{1}{\alpha_2}.
\end{equation}
Let's now consider a slightly more complicated example.

\paragraph{Example 3.2}
We now consider a basis given by the element,

\begin{equation}
    \phi_1 = d\ln\left(\frac{x-1}{x^{-1}}\right)
\end{equation}
where we have indicated the hyperplane at infinity by $x^{-1}$. The self intersection number of this form gives

\begin{equation}\label{eq3.17}
    \langle{\phi_1,\phi_1\rangle}^{-1} = \left(\frac{1}{\alpha_2}-\frac{1}{\alpha_1+\alpha_2}\right)^{-1} = \frac{(\alpha_1+\alpha_2)\alpha_2}{\alpha_1}.
\end{equation}
We have the partial intersection number

\begin{equation}\label{eq3.18}
    \langle{\varphi_{\A_1},\phi_1\rangle} = \frac{1}{\alpha_2}.
\end{equation}
using which the twisted Riemann period relation becomes

\begin{equation}
    \langle{\varphi_{\A_1},\phi_1\rangle} \langle{\phi_1,\phi_1\rangle}^{-1} \langle{\varphi_{\A_1},\phi_1\rangle} = \frac{1}{\alpha_1} + \frac{1}{\alpha_2}.
\end{equation}

Before considering one final example, we note that it is quite easy to see how the KLT matrix in equation (\ref{eq3.17}) fails to be planar - indeed, it isn't really even the inverse of an amplitude most of the time! Consider the case of $6$-particle scattering in $\phi^4$ theory, in which we choose $\alpha_1 = X_{14}$ and $\alpha_2 = X_{36}$. Clearly, $X_{14}+X_{36}$ is not a Mandelstam variable; it doesn't even correspond to a scattering channel. 

\paragraph{Example 3.3}
Let's consider a final example, now using a mixed basis. On the left, we consider the basis given by $\phi_1 = \varphi_{\A_1}$ and on the right we consider the basis given by $\phi'_2=\varphi^{(2)}_{\A_1}$. This leads to the twisted period relation,

\begin{equation}
    \langle{\varphi_{\A_1},\varphi_{\A_1}\rangle} = \langle{\varphi_{\A_1},\phi_1\rangle}\langle{\phi_1',\phi_1\rangle}^{-1}\langle{\phi_1',\varphi_{\A_1}\rangle},
\end{equation}
which can be seen to be true essentially trivially, but can be verified readily by making use of equations (\ref{eq3.12}) and (\ref{eq3.18}).

\subsection{Two-Dimensional KLT Relations}
Let us now discuss the case of KLT relations that arise out of the twisted Riemann period relations for two dimensional accordiohedra intersecting. We have two possibilities, the square and pentagon accordiohedra. 

A square is defined by the hyperplane arrangement

\begin{equation}
\begin{aligned}
    f_{1} &= 1-y \\
    f_{2} &= 1-x \\
    f_{3} &= 1+y \\
    f_{4} &= 1+x.
\end{aligned}
\end{equation}
With these we have the complement and twist as

\begin{equation}
    \A_{2}^{(1)} = \mathbb{CP}^{2}-\lbrace{f_1,f_2,f_3,f_4\rbrace}
\end{equation}

\begin{equation}
    \omega_{\A^{(1)}_{2}} = \beta_{1}d\ln f_{1} + \beta_{2}d\ln f_{2} + \beta_{3}d\ln f_{3} + \beta_{4}d\ln f_{4}. 
\end{equation}

Similarly for the case of the pentagon we use the data

\begin{equation}
    \begin{aligned}
    g_{1} &= 1-x\\
    g_{2} &= 1+y\\
    g_{3} &= 2+x\\
    g_{4} &= 2+x-y\\
    g_{5} &= 2-y,
    \end{aligned}
\end{equation}

\begin{equation}
    \A_{2}^{(2)} = \mathbb{CP}^{2}-\lbrace{g_1,g_2,g_3,g_4,g_{5}\rbrace},
\end{equation}
and
\begin{equation}
    \omega_{\A^{(2)}_{2}} = \gamma_{1}d\ln g_{1} + \gamma_{2}d\ln g_{2} + \gamma_{3}d\ln g_{3} + \gamma_{4}d\ln g_{4} + \gamma_{5}d\ln g_{5}. 
\end{equation}

Consider first the case of the square. Take the form

\begin{equation}
    \varphi_{\A^{(1)}_{2}} = d\ln\left(\frac{f_{1}}{f_{3}}\right)\wedge d\ln\left(\frac{f_{2}}{f_{4}}\right).
\end{equation}
The self intersection number of this form on $A^{(1)}_{2}$ is computed as

\begin{equation}
    \langle{\varphi_{\A^{(1)}_{2}},\varphi_{\A^{(1)}_{2}}\rangle} = \left(\frac{1}{\beta_{1}}+\frac{1}{\beta_3}\right)\left(\frac{1}{\beta_{2}}+\frac{1}{\beta_4}\right).
\end{equation}
Indeed, this is just the square of the amplitude we obtained from the one dimensional case. Accordingly, we have the following.

\paragraph{Example 3.4}
Since the twisted cohomology group in this case is one dimensional, we need to specify again only one form in order to derive the KLT relations. The KLT relations are determined here entirely by the KLT relation for the line. For purposes of clarity, suppose we consider the basis given by

\begin{equation}
    \phi_{1} = d\ln\left(\frac{f_{1}}{y^{-1}}\right)\wedge d\ln\left(\frac{f_{2}}{x^{-1}}\right).
\end{equation}
The KLT relation

\begin{equation}
    \langle{\varphi_{\A^{(1)}_{2}},\varphi_{\A^{(1)}_{2}}\rangle} = \langle{\varphi_{\A^{(1)}_{2}},\phi_{1}\rangle} \langle{\phi_{1},\phi_{1}\rangle}^{-1} \langle{\phi_{1},\varphi_{\A^{(1)}_{2}}\rangle}
\end{equation}
can be established by noting that

\begin{equation}
    \langle{\varphi_{\A^{(1)}_{2}},\phi_{1}\rangle} = \frac{1}{\beta_{1}\beta_{2}}
\end{equation}
and

\begin{equation}
    \langle{\phi_{1},\phi_{1}\rangle} = \left(\frac{1}{\beta_{1}}-\frac{1}{\beta_1 + \beta_3}\right)\left(\frac{1}{\beta_{2}}-\frac{1}{\beta_2 + \beta_4}\right).
\end{equation}

\begin{figure}[H]
\centering
\includegraphics[width=0.59\textwidth]{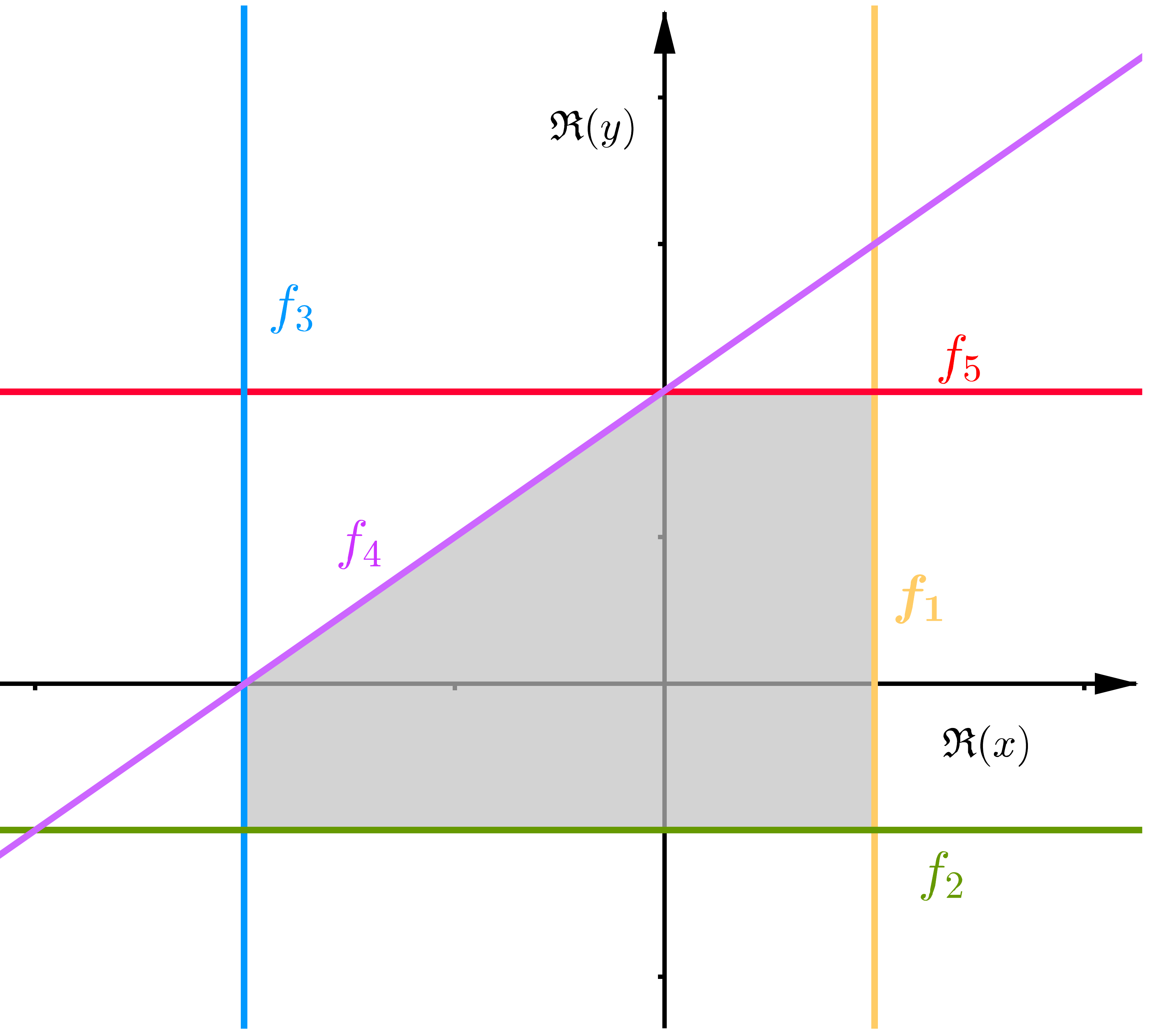}
\caption{Real Slice of Arrangement $\A^{(2)}_{1}$}\label{fig3.1}
\end{figure}

Moving on, we now consider three examples for the pentagon. Defining the form

\begin{equation}
    \varphi_{\A^{(2)}_{2}} = d\ln\left(\frac{g_{1}}{g_{2}}\right)\wedge d\ln\left(\frac{g_{2}}{g_{3}}\right) + d\ln\left(\frac{g_{3}}{g_{4}}\right)\wedge d\ln\left(\frac{g_{4}}{g_{5}}\right) + d\ln\left(\frac{g_{5}}{g_{1}}\right)\wedge d\ln\left(\frac{g_{1}}{g_{3}}\right)
\end{equation}
the self intersection number is

\begin{equation}\label{3.41}
    \langle{\varphi_{\A^{(2)}_{2}},\varphi_{\A^{(2)}_{2}}\rangle} = \frac{1}{\gamma_{1}\gamma_{2}} + \frac{1}{\gamma_{2}\gamma_{3}} + \frac{1}{\gamma_{3}\gamma_{4}} + \frac{1}{\gamma_{4}\gamma_{5}} + \frac{1}{\gamma_{5}\gamma_{1}}.
\end{equation}
To determine a basis for the cohomology group, we note that the dimension of the group is $4$, meaning that we need to specify four basis elements. We can glean a natural way of doing this by inspecting a real slice of the hyperplane arrangement, as shown in figure \ref{fig3.1} \cite{Kalyanapuram:2020tsr}.

One can infer simply by inspection that there are four bounded chambers, which is a simple consistency check of the dimension of the cohomology group. An element of the cohomology group is furnished by a chamber; this can be seen by noting the Poincar\'e duality - a form with singularities on the boundaries of a chamber will be dual to the chamber, a cycle on the complement space. 

\paragraph{Example 3.5}
Consider a basis for the cohomology group given by the four \emph{distinct} bounded chambers. These are given by

\begin{equation}
    \phi_{1} =  d\ln\left(\frac{g_{1}}{g_{2}}\right)\wedge d\ln\left(\frac{g_{2}}{g_{3}}\right) + d\ln\left(\frac{g_{3}}{g_{4}}\right)\wedge d\ln\left(\frac{g_{4}}{g_{5}}\right) + d\ln\left(\frac{g_{5}}{g_{1}}\right)\wedge d\ln\left(\frac{g_{1}}{g_{3}}\right),
\end{equation}

\begin{equation}
    \phi_{2} = d\ln\left(\frac{g_{2}}{g_{3}}\right)\wedge d\ln\left(\frac{g_{3}}{g_{4}}\right),
\end{equation}

\begin{equation}
    \phi_{3} = d\ln\left(\frac{g_{3}}{g_{4}}\right)\wedge d\ln\left(\frac{g_{4}}{g_{5}}\right)
\end{equation}
and

\begin{equation}
    \phi_{4} = d\ln\left(\frac{g_{4}}{g_{5}}\right)\wedge d\ln\left(\frac{g_{5}}{g_{1}}\right).
\end{equation}
To determine the KLT relations, we have to compute the partial amplitudes $\langle{\varphi_{\A^{(2)}_{1}},\phi_{i}\rangle}$ and the matrix $\langle{\phi_i,\phi_j\rangle}$. We have

\begin{equation}\label{eq3.39}
    \begin{split}
    &\langle{\varphi_{\A^{(2)}_{1}},\phi_{1}\rangle} = \frac{1}{\gamma_{1}\gamma_{2}} + \frac{1}{\gamma_{2}\gamma_{3}} + \frac{1}{\gamma_{3}\gamma_{4}} + \frac{1}{\gamma_{4}\gamma_{5}} + \frac{1}{\gamma_{5}\gamma_{1}} \\
    &\langle{\varphi_{\A^{(2)}_{1}},\phi_{3}\rangle} = \frac{1}{\gamma_{3}\gamma_{4}}+\frac{1}{\gamma_{4}\gamma_{5}}
   \end{split}\qquad
   \begin{split}
       &\langle{\varphi_{\A^{(2)}_{1}},\phi_{2}\rangle} = \frac{1}{\gamma_{2}\gamma_{3}} + \frac{1}{\gamma_{3}\gamma_{4}} \\
    &\langle{\varphi_{\A^{(2)}_{1}},\phi_{4}\rangle} = \frac{1}{\gamma_4\gamma_5}+\frac{1}{\gamma_5\gamma_1}.
   \end{split}
\end{equation}
and

\begin{equation}
      \begin{split}
    &\langle{\phi_{1},\phi_{1}\rangle} = \frac{1}{\g_{1}\g_{2}} + \frac{1}{\g_{2}\g_{3}} + \frac{1}{\g_{3}\g_{4}} + \frac{1}{\g_{4}\g_{5}} + \frac{1}{\g_{5}\g_{1}} \\
    &\langle{\phi_{1},\phi_{2}\rangle} = \frac{1}{{\g_{4}\g_{5}}}+\frac{1}{{\g_{5}\gamma_{1}}}\\
    &\langle{\phi_{1},\phi_{3}\rangle} = \frac{1}{{\g_{3}\g_{4}}}+\frac{1}{{\g_{4}\g_{5}}} \\
    &\langle{\phi_{1},\phi_{4}\rangle} = \frac{1}{{\g_{2}\g_{3}}}+\frac{1}{{\g_{3}\g_{4}}}\\
    &\langle{\phi_{2},\phi_{2}\rangle} = \frac{1}{\g_{2}\g_{3}} + \frac{1}{\g_{3}\g_{4}} + \frac{1}{\g_{4}\g_{2}}\\
   \end{split}\qquad
   \begin{split}
    &\langle{\phi_{2},\phi_{3}\rangle} = \frac{1}{{\g_{4}\g_{5}}}  \\
    &\langle{\phi_{2},\phi_{4}\rangle} = 0\\
    &\langle{\phi_{3},\phi_{3}\rangle} = \frac{1}{\g_{3}\g_{4}} + \frac{1}{\g_{4}\g_{5}} + \frac{1}{\g_{5}\g_{3}}\\
    &\langle{\phi_{3},\phi_{4}\rangle} = \frac{1}{{\g_{3}\g_{4}}} \\
    &\langle{\phi_{4},\phi_{4}\rangle} = \frac{1}{\g_{4}\g_{5}} + \frac{1}{\g_{5}\g_{1}} + \frac{1}{\g_{1}\g_{4}}\\
   \end{split}
\end{equation}
With these, the KLT relation is given by the resolution,

\begin{equation}
    \langle{\varphi_{\A^{(2)}_{1}},\varphi_{\A^{(2)}_{1}}\rangle} = \langle{\varphi_{\A^{(2)}_{1}},\phi_{i}\rangle}\langle{\phi_{j},\phi_{i}\rangle}^{-1}\langle{\phi_{j},\varphi_{\A^{(2)}_{1}}\rangle}
\end{equation}
which has been verified using {\scshape Mathematica}.

The expressions for the inverse of the KLT matrix $\langle{\phi_i,\phi_j\rangle}$ computed in {\scshape Mathematica} were extremely unilluminating and baroque. As a result of this, we have not attempted to reproduce the results herein. It may be of value to carry out a systematic study of this issue in future work.

\paragraph{Example 3.6}
Consider now the following basis for the cohomology group of the pentagon

\begin{equation}
    \phi'_{1} =  d\ln\left(\frac{g_{2}}{g_{3}}\right)\wedge d\ln\left(\frac{g_{3}}{g_{4}}\right),
\end{equation}

\begin{equation}
    \phi'_{2} = d\ln\left(\frac{g_{2}}{g_{1}}\right)\wedge d\ln\left(\frac{g_{1}}{g_{4}}\right),
\end{equation}

\begin{equation}
    \phi'_{3} = d\ln\left(\frac{g_{4}}{g_{5}}\right)\wedge d\ln\left(\frac{g_{5}}{g_{1}}\right)
\end{equation}
and

\begin{equation}
    \phi'_{4} = d\ln\left(\frac{g_{3}}{g_{4}}\right)\wedge d\ln\left(\frac{g_{4}}{g_{5}}\right).
\end{equation}
Now, we have to compute $\langle{\varphi_{\A^{(2)}_{1}},\phi'_{i}\rangle}$ and the matrix $\langle{\phi_i',\phi'_j\rangle}$. We have,

\begin{equation}\label{3.46}
    \begin{split}
    &\langle{\varphi_{\A^{(2)}_{1}},\phi'_{1}\rangle} = \frac{1}{\gamma_{2}\gamma_{3}} + \frac{1}{\gamma_{3}\gamma_{4}}\\
    &\langle{\varphi_{\A^{(2)}_{1}},\phi'_{3}\rangle} =  \frac{1}{\gamma_4\gamma_5}+\frac{1}{\gamma_5\gamma_1}
   \end{split}\qquad
   \begin{split}
       &\langle{\varphi_{\A^{(2)}_{1}},\phi'_{2}\rangle} = -\frac{1}{\gamma_{1}\gamma_{2}}\\
    &\langle{\varphi_{\A^{(2)}_{1}},\phi'_{4}\rangle} = \frac{1}{\gamma_{3}\gamma_{4}}+\frac{1}{\gamma_{4}\gamma_{5}},
   \end{split}
\end{equation}
and
\begin{equation}
      \begin{split}
    &\langle{\phi'_{1},\phi'_{1}\rangle} = \frac{1}{\g_{2}\g_{3}} + \frac{1}{\g_{3}\g_{4}} + \frac{1}{\g_{2}\g_{4}}\\
    &\langle{\phi'_{1},\phi'_{2}\rangle} = \frac{1}{\g_{2}\g_{4}}\\
    &\langle{\phi'_{1},\phi'_{3}\rangle} =  0\\
    &\langle{\phi'_{1},\phi'_{4}\rangle} = \frac{1}{\g_{3}\g_{4}}\\
    &\langle{\phi'_{2},\phi'_{2}\rangle} = \frac{1}{\g_{1}\g_{2}} + \frac{1}{\g_{1}\g_{4}} + \frac{1}{\g_{2}\g_{4}}\\
   \end{split}\qquad
   \begin{split}
    &\langle{\phi'_{2},\phi'_{3}\rangle} =  \frac{1}{\g_{1}\g_{4}} \\
    &\langle{\phi'_{2},\phi'_{4}\rangle} = 0\\
    &\langle{\phi'_{3},\phi'_{3}\rangle} = \frac{1}{\g_{4}\g_{5}} + \frac{1}{\g_{5}\g_{1}} + \frac{1}{\g_{1}\g_{4}}\\
    &\langle{\phi'_{3},\phi'_{4}\rangle} = \frac{1}{\g_{4}\g_{5}}\\
    &\langle{\phi'_{4},\phi'_{4}\rangle} = \frac{1}{\g_{3}\g_{4}} + \frac{1}{\g_{4}\g_{5}} + \frac{1}{\g_{3}\g_{5}}.\\
   \end{split}
\end{equation}
Using {\scshape Mathematica}, we have verified that the KLT relation

\begin{equation}
    \langle{\varphi_{\A^{(2)}_{1}},\varphi_{\A^{(2)}_{1}}\rangle} = \langle{\varphi_{\A^{(2)}_{1}},\phi'_{i}\rangle}\langle{\phi'_{j},\phi'_{i}\rangle}^{-1}\langle{\phi'_{j},\varphi_{\A^{(2)}_{1}}\rangle}
\end{equation}
holds.

\paragraph{Example 3.7}
Finally, in this example, we consider a case of the mixed KLT relation for the pentagon, by inserting two different bases in the intersection number (\ref{3.41}). We use the bases $\lbrace{\phi_{i}\rbrace}$ and $\lbrace{\phi'_{i}\rbrace}$ and the resolution,

\begin{equation}\label{eq3.49}
     \langle{\varphi_{\A^{(2)}_{1}},\phi'_{j}\rangle}\langle{\phi'_{i},\phi_{j}\rangle}^{-1}\langle{\phi_{i},\varphi_{\A^{(2)}_{1}}\rangle}.
\end{equation}
Proving the KLT relation in this case amounts to proving that this quantity is equal to (\ref{3.41}). To do this, we have the quantities $\langle{\phi_{i},\varphi_{\A^{(2)}_{1}}\rangle}$ and $\langle{\varphi_{\A^{(2)}_{1}},\phi'_{j}\rangle}$ already computed. We are now left with the task of finding the elements of the matrix $\langle{\phi_{i},\phi'_{j}\rangle}$. We have

\begin{equation}
    \begin{split}
    &\langle{\phi_{1},\phi'_{1}\rangle} = \frac{1}{\g_{2}\g_{3}}+\frac{1}{\g_{3}\g_{4}} \\
    &\langle{\phi_{1},\phi'_{2}\rangle} = -\frac{1}{\g_{1}\g_{2}}\\
    &\langle{\phi_{1},\phi'_{3}\rangle} = \frac{1}{\g_{4}\g_{5}}+ \frac{1}{\g_{5}\g_{1}}\\
    &\langle{\phi_{1},\phi'_{4}\rangle} = \frac{1}{\g_{3}\g_{4}}+\frac{1}{\g_{4}\g_{5}}\\
    &\langle{\phi_{2},\phi'_{1}\rangle} = \frac{1}{\g_{2}\g_{3}} + \frac{1}{\g_{2}\g_{4}}\\
    &\langle{\phi_{2},\phi'_{2}\rangle} =  \frac{1}{\g_{2}\g_{4}} \\
    &\langle{\phi_{2},\phi'_{3}\rangle} = 0\\
    &\langle{\phi_{2},\phi'_{4}\rangle} = \frac{1}{\g_{3}\g_{d}}\\
   \end{split}\qquad
   \begin{split}
    &\langle{\phi_{3},\phi'_{1}\rangle} = \frac{1}{\g_{3}\g_{4}}\\
    &\langle{\phi_{3},\phi'_{2}\rangle} = 0\\
    &\langle{\phi_{3},\phi'_{3}\rangle} = \frac{1}{\g_{4}\g_{5}}\\
    &\langle{\phi_{3},\phi'_{4}\rangle} = \frac{1}{\g_{3}\g_{4}}+\frac{1}{\g_{4}\g_{5}}+\frac{1}{\g_{5}\g_{3}}\\
    &\langle{\phi_{4},\phi'_{1}\rangle} = 0\\
    &\langle{\phi_{4},\phi'_{2}\rangle} = \frac{1}{\g_{1}\g_{4}}\\
    &\langle{\phi_{4},\phi'_{3}\rangle} = \frac{1}{\g_{4}\g_{5}}+\frac{1}{\g_{5}\g_{1}}+\frac{1}{\g_{1}\g_{4}}\\
    &\langle{\phi_{4},\phi'_{4}\rangle} = \frac{1}{\g_{4}\g_{5}}.\\
   \end{split}
\end{equation}
Using this, (\ref{eq3.39}) and (\ref{3.46}), we have verified using {\scshape Mathematica} that the KLT relation (\ref{eq3.49}) holds.

As a final note, we point out the challenge involved in extending this framework to higher points. Take the case of a $10$-point amplitude in $\phi^4$. Neglecting the trivial cases of the box and product polytope, we saw that the dimension of the cohomology class of the associahedron type polytope was $24$ dimensional while the cohomology class of the Lucas type polytope was $12$ dimensional. The three dimensional nature of these polytopes makes a direct counting and enumeration of the bounded chambers more complicated. One possible approach to writing out cohomology classes would perhaps be to find the intersection lattice of the arrangement and use that data to extract information about chambers, which could serve as an interesting direction of future research.

\pagebreak

\section{Conclusion}\label{sec:conclusion}
In this article, we have studied the implications of the twisted Riemann period relations to the study of accordiohedra. We made use of the recent analysis by the author \cite{Kalyanapuram:2019nnf,Kalyanapuram:2020tsr} which revealed an interpretation of scattering amplitudes at tree level in scalar theories are patterns of intersection between accordiohedra. The fact that these amplitudes could be written as intersection numbers naturally suggested that the twisted Riemann period relations may be an interesting direction of study, which is what we have set out to carry out in the note.

The twisted Riemann period relations have already been employed in the study of theories with cubic interactions, where they were shown to give rise to the KLT relations between gravity and Yang-Mills amplitudes at tree level. This allows us to make some comments on the relationship borne by our analysis to these previous studies as well as lay out some directions for further study. 

In Section \ref{sec:scattering}, we have presented a synthesis of the data derived so far in the literature that describe the combinatorial structures of accordiohedra for a number of examples. We treated examples of accordiohedra in $\phi^4$ theory as well as the theory $\phi^3+\phi^4$ with mixed vertices. After providing details on how these can be understood as hyperplane arrangements in $\mathbb{CP}^{n}$, we used the methods of twisted cohomology theory to derive the analogue of the scattering equations due to Cachazo, He and Yuan for these more generic quantum field theories. A salient feature throughout was the fact that since for a given scattering process accordiohedra are generically not unique, we obtain a \emph{family} of scattering equations, rather than a single one as in the case of the CHY formalism. 

We then made use of the fact that the number of equations of the scattering equations coincides with the dimension of the cohomology group to compute the dimensions of all the relevant groups in our analysis. Applying now the fact that scattering amplitudes of the relevant scalar theories are simply intersection numbers of the accordiohedra, in Section \ref{sec:relations} we found the KLT relations for these theories by studying the twisted Riemann period relations, which amount to simply an insertion of two complete sets of cohomology bases into the intersection number. Again, an important feature of the analysis was the lack of uniqueness; one KLT relation is present for every accordiohedron contributing to the scattering. 

There are a few key differences that distinguish the KLT relations in this work from the original relations between gravity and Yang-Mills amplitudes. In the original KLT relations, the KLT matrix was a very special object from the point of view of quantum field theory - it is the inverse of the matrix of partial amplitudes in the biadjoint scalar theory. No such simple picture seems to arise in our analysis. Indeed, certain bases give rise to elements of the KLT matrix containing poles which are spurious and cannot arise out of any scattering channel\footnote{Some of the terms in the generalized KLT matrix are noncausal in that they violate the Steinmann relations. They cannot arise from a consistent quantum field theory. I thank Simon Caron-Huot for pointing this out.}. Another important difference is more practical in bent. In this work, we have focused on the simplest examples, dealing only with one and two dimensional examples. Other than the obvious ease of computation this facilitates, we have saved ourselves of having to deal with the complicated geometry of hyperplanes in higher dimensions. We have been able to specify cohomology bases essentially by inspection. It would be quite unsatisfactory if one were forced to do this at higher dimensions as well. Accordingly, one definite line of future research could be to study the possibility of finding a canonical presentation of cohomology classes for accordiohedron hyperplane arrangements\footnote{A possible answer to the question posed here might be to refine the basis due to Aomoto described in \cite{Abe:2015ucn,Abe:2018bgl}. It doesn't seem to always apply in the case by accordiohedra, due in part to the specific nature of the arrangement.}.

Field theory amplitudes are not alone in admitting a representation as intersection numbers. String amplitudes in the closed and open sectors as well as more exotic amplitude with $\alpha'$ corrections such as $Z$-theory \cite{Mafra:2011nv,Mafra:2011nw,Carrasco:2016ldy,Carrasco:2016ygv,Mafra:2016mcc} can also be represented as intersection numbers. There also exist deformations of the Parke-Taylor factor \cite{Stieberger:2013hza,Stieberger:2013nha,Mizera:2017sen} which recast string amplitudes as essentially CHY type formulae. There is a natural generalization of the Koba-Nielsen factor in the accordiohedron context, which was touched upon briefly in \cite{Kalyanapuram:2020vil} as well as in more conventional string-like contexts \cite{Baadsgaard:2015hia,Baadsgaard:2015ifa,Baadsgaard:2015voa,Baadsgaard:2016fel}. Exploring the analogue of $\alpha'$ corrections in generic scalar theories and the possibilities of double copy relations in terms of string-like generalizations of scalar amplitudes studied in \cite{Mizera:2016jhj,Kalyanapuram:2019nnf} might be interesting.

Another very popular avatar of the double copy is due to Bern, Carrasco and Johansson (see \cite{Bern:2008qj,Bern:2010tq,Bern:2010ue,Bern:2019prr} and references thereof and therein), in which the double copy is carried out at the level of individual Feynman diagrms, where kinematical and colour factors in the numerators obery identities that enable an essential 'squaring' of a gluon amplitude into a gravity amplitude. It has transpired of late that string amplitudes and intersection theory \cite{Mizera:2019gea} may play a role in providing some insight into their origins. It may prove worthwhile to pursue this connection for theories with higher point interactions using the tools developed here.

In this work and work on accordiohedra more generally, we have not yet tried to understand the implications of the theory of polytopes to the study of quantum field theory with numerators and possible double copy structures therein. One may consider extending the analysis carried out in the work to the study of theories such as $\mathcal{N}=4$ super Yang-Mills and $\mathcal{N}=8$ supergravity, which admit geometric representations \cite{ArkaniHamed:2008gz,ArkaniHamed:2009dn,ArkaniHamed:2009si,ArkaniHamed:2009vw,ArkaniHamed:2010gg,ArkaniHamed:2010gh,ArkaniHamed:2010kv,ArkaniHamed:2012nw,Arkani-Hamed:2013jha,Arkani-Hamed:2014dca} as well. We note here parenthetically that the construction of double copy relations for polynomial vertices may also make possible future analyses of double copy relations without having recourse to string theory, thereby possibily enabling future applications of double copy relations to realistic theories.

Finally, we note that the intersection theory of accordiohedra has been developed thus far with no recourse being had to string theory methods. However, higher point interactions have been studied using string based methods in the past \cite{Baadsgaard:2016fel}. Modern developments in superstring field theory  \cite{Moosavian:2017fta,Moosavian:2017qsp,Moosavian:2017sev,Moosavian:2019ydz}, especially in light of the use made of hyperbolic geometry in the closed sector, may consequently prove to be valuable in decoding the relationship between higher loop interactions and polytopes and in shedding light on the nature of the double copy more generally.

\pagebreak

\appendix

\titleformat{name=\section}[display]
{\normalfont}
{\footnotesize\textsc{Appendix \thesection}}
{0pt}
{\Large\bfseries}
[\vspace{-10pt}\color{lapis}\rule{\textwidth}{2pt}]

\bibliographystyle{JHEP}
\bibliography{v1}

\end{document}